\documentclass{article}
\usepackage{epsfig}
\usepackage{color}
\usepackage{amsmath}
\usepackage{amssymb}
\usepackage{lscape}

\makeatletter
\@addtoreset{equation}{section}

\makeatother

\begin{document}

%\fontsize{14pt}{0.8cm}\selectfont
\fontsize{11pt}{0.5cm}\selectfont

\begin{center}

\vspace*{4cm}
{
\huge

Differential Equations for Wilson Loops 

\vspace{5mm}

in ABJM Theory
}

\vspace{2cm}

{
\Large 
Takao Suyama\footnote{suyama.takao@nihon-u.ac.jp}
}

\vspace{1cm}

{
\Large

College of Engineering, Nihon University, 

\vspace{5mm}

Fukushima 963-8642, Japan
}

\vspace{2cm}

{\bf Abstract}
\end{center}

\vspace{5mm}

We derive a system of differential equations which are satisfied by the vevs of BPS Wilson loops and 't~Hooft coupling of ABJM theory. 
They are Picard-Fuchs equations of an algebraic curve defined by the derivative of the planar resolvent of the corresponding matrix model. 
The weak and strong coupling behaviors can be reproduced by their local solutions around regular singularities. 
We also obtain a recursion relation which can be used to determine the planar vevs of BPS Wilson loops in arbitrary representations. 

\newpage

\tableofcontents

\vspace{1cm}

\section{Introduction}

\vspace{5mm}

ABJM theory was proposed in \cite{Aharony:2008ug} as an example of the boundary CFT for AdS/CFT correspondence \cite{Maldacena:1997re}. 
This is an ${\cal N}=6$ superconformal Chern-Simons theory coupled to matter fields in three-dimensions. 
The gauge group is ${\rm U}(N)\times{\rm U}(N)$, and the levels of the first and the second unitary factors are $k$ and $-k$, respectively. 
The theory can be extended to the one with the gauge group ${\rm U}(M)\times{\rm U}(N)$. 
This is often called ABJ theory, and it is extensively studied in \cite{Aharony:2008gk}. 
Note that the Lagrangian of ABJ theory had already appeared previously in \cite{Hosomichi:2008jb}, based on the construction of \cite{Gaiotto:2008sd}. 

Soon after the appearance of ABJM theory, Wilson loops which preserves some amount of supersymmetry were constructed. 
The first example of the BPS Wilson loops preserve 1/6 of the supersymmetry \cite{Drukker:2008zx}\cite{Chen:2008bp}\cite{Rey:2008bh}, which is a natural extension of the 1/2 BPS Wilson loop \cite{Rey:1998ik}\cite{Maldacena:1998im} in ${\cal N}=4$ super-Yang Mills theory in four dimensions. 
The perturbative calculation of its vev in the 't~Hooft coupling $\lambda:=N/k$ was also performed, resulting in 
\begin{equation}
W\ =\ 1+\frac56\pi^2\lambda^2+{\cal O}(\lambda^4). 
\end{equation} 
This calculation was extended to higher orders in \cite{Bianchi:2016yzj}. 
The second example is the 1/2 BPS Wilson loop constructed in \cite{Drukker:2009hy}. 
This is constructed using also fermionic fields. 
The structure is more complicated than the 1/6 BPS Wilson loops, and it is understood in terms of a supergroup \cite{Drukker:2009hy}\cite{Lee:2010hk}. 
There is a review \cite{Drukker:2019bev} on Wilson loops in theories including ABJM theory. 

The investigation of ABJM theory was accelerated by the localization formulas obtained in \cite{Kapustin:2009kz}. 
The resulting expression for the partition function resembles the one for matrix models, and therefore the matrix model techniques were applied to ABJM theory \cite{Suyama:2009pd}\cite{Marino:2009jd}\cite{Drukker:2010nc}. 
Remarkably, it was noticed in \cite{Marino:2009jd} that there is a relation between the localized partition function and a topological string theory, through a matrix model of \cite{Halmagyi:2003ze}, which is then used to analyze the strong coupling limit of the BPS Wilson loops and the free energy \cite{Marino:2009jd}\cite{Drukker:2010nc}. 
The leading behavior of Wilson loops for large $\lambda$ turns out to be 
\begin{equation}
W\ \sim\ e^{\pi\sqrt{2\lambda}}, 
   \label{gravity dual}
\end{equation}
which reproduces the dual gravity calculation \cite{Drukker:2008zx}\cite{Chen:2008bp}\cite{Rey:2008bh} for 1/2 BPS case\footnote{
The gravity dual of the 1/6 BPS Wilson loops were conjectured \cite{Drukker:2008zx}\cite{Rey:2008bh} to be a smeared string which gives the same behavior as (\ref{gravity dual}). 
Later, a 1/6 BPS classical string solution was constructed in \cite{Aguilera-Damia:2014qgy}, but it turned out that it corresponds to a different Wilson loop operator. 
}. 
The 't~Hooft coupling $\lambda$ and the free energy are determined as a solution of the Picard-Fuchs equations related to the background geometry of the topological string theory. 

There is another powerful machinery for analyzing ABJM theory and some other Chern-Simons-matter theories, known as the Fermi gas formalism, introduced in \cite{Marino:2011eh}. 
This machinery is so powerful that even non-perturbative structure of the theory can be analyzed precisely. 
The remarkable results on the free energy of ABJM theory are summarized in the reviews \cite{Hatsuda:2015gca}\cite{Marino:2016new}. 
The Fermi gas formalism can also be applied to the BPS Wilson loops \cite{Klemm:2012ii}\cite{Hatsuda:2013yua}. 

The relation between ABJM theory and the topological string theory has achieved remarkable success so far, so it is naturally expected that similar relations might also be available in other Chern-Simons-matter theories. 
Recently, it is conjectured in \cite{Grassi:2014zfa} that topological string theories can be related systematically to some quantum mechanical systems including ones appearing in the Fermi gas formalism. 
This is called TS/ST correspondence. 
See a review \cite{Marino:2015nla}. 
However, it seems that the existence of a general relation to Chern-Simons-matter theories is still unclear. 

\vspace{5mm}

In this paper, we derive a system of differential equations satisfied by the BPS Wilson loops and the 't~Hooft coupling of ABJM theory. 
This is achieved by considering an algebraic curve defined by the derivative of the resolvent of a matrix model for ABJM theory. 
It was observed in \cite{Suyama:2016nap} that the derivative of the resolvent is much simpler than the resolvent itself. 
Note that the curve in this paper is different from the one discussed in \cite{Marino:2009jd}. 
The Wilson loops and the 't~Hooft coupling can be written as integrals along a closed contour on the curve. 
This enables us to obtain the Picard-Fuchs equations for them. 
We analyze the solution of the equations for weak and strong coupling limit, and reproduce the known results. 
As a by-product, we find a recursion relation for quantities defined as in (\ref{def-pi_n}) below, which can be used to obtain vevs of BPS Wilson loops in higher representations in the planar limit. 
Note that Picard-Fuchs equations are also employed in the context of Seiberg-Witten theory. 
See for example \cite{Isidro:2000zw}. 

Our analysis is confined in the planar limit, so we cannot add any new findings to the existing results which are far more precise and deep, like the ones in \cite{Hatsuda:2015gca}\cite{Marino:2016new}. 
Instead, our aim is to explore Chern-Simons-matter theories as general as possible. 
Since our derivation of the Picard-Fuchs equations does not rely on topological string theory, we expect that it can be applied to more general theories even though there is no known relation to string theory. 
The investigation on general Chern-Simons-matter theories would provide more insights into the presence/absence of the gravity dual, for example, which would characterize the family of theories with gravity duals. 
The shape of a thing is determined by its boundary. 

This paper is organized as follows. 
In section \ref{pureCS}, to illustrate our analysis, we consider pure Chern-Simons theory. 
For this theory, the vev of the Wilson loop is known exactly \cite{Witten:1988hf}. 
After a short summary on the matrix model of pure Chern-Simons theory in subsection \ref{Planar limit}, we derive the Picard-Fuchs equations in subsection \ref{PF-pureCS}. 
The equation for the Wilson loop can be solved easily. 
The solution is a function of a parameter which is related to the 't~Hooft coupling given by solving another Picard-Fuchs equation. 
In subsection \ref{Recursion relation}, we also obtain the recursion relation mentioned above, whose solution is given as a generating function. 
Then, we analyze ABJM theory in section \ref{ABJM theory}. 
The derivation of the Picard-Fuchs equations for the Wilson loops and the 't~Hooft coupling for ABJM theory in subsection \ref{PF-ABJM} is parallel to the one for pure Chern-Simons theory, but it is more complicated. 
As a result, we find that the Wilson loops and the 't~Hooft coupling are given in terms of solutions of (inhomogeneous) Heun equations. 
The local solutions around regular singularities of the equations reproduce the known weak and strong coupling expansions of the Wilson loop. 
The recursion relation for ABJM theory is obtained in subsection \ref{recursion relation ABJM} which turns out to be a rather simple 5-term relation. 
Section \ref{Discussion} is devoted to discussion. 
In appendix \ref{curve-derivation}, we summarize the derivation of the curves, on which our analysis is based, purely within the matrix model context. 
In appendix \ref{integral estimate}, we estimate the behavior of the 't~Hooft coupling in the parameter region corresponding to the strong coupling. 

\vspace{1cm}

\section{Pure Chern-Simons theory}
   \label{pureCS}

\vspace{5mm}

The quantum field theory for the Chern-Simons action with a non-Abelian gauge group was defined in \cite{Witten:1988hf}. 
We call this pure Chern-Simons theory. 
This is a three-dimensional gauge theory defined on a manifold $M$. 
This theory is independent of the metric on $M$, and therefore it is a topological field theory. 
Indeed, the Chern-Simons action is written without using the metric. 
A metric dependence may arise from the gauge fixing, but it was shown in \cite{Witten:1988hf} that it can be canceled by a counterterm. 

The partition function of pure Chern-Simons theory can be calculated exactly by using the topological properties of the theory. 
The important observables of this theory is the Wilson loops. 
The correlation functions of the Wilson loops can also be calculated, and they give topological invariants of knots and links in $M$. 
For example, when the gauge group is ${\rm SU}(N)$ and the level is $k$, 
the vev of the Wilson loop in the fundamental representation for an unknotted circle in $S^3$ is given as 
\begin{equation}
\langle W \rangle\ =\ \frac{q^{N/2}-q^{-N/2}}{q^{1/2}-q^{-1/2}}, \hspace{1cm} q\ :=\ e^{2\pi i/(k+N)}. 
   \label{Witten-WL}
\end{equation}

It was found in \cite{Marino:2002fk} that the exact expression for the partition function can be rewritten as a finite-dimensional integral. 
For the case when the gauge group is ${\rm U}(N)$ and the manifold $M$ is $S^3$, the resulting integral expression is proportional to (\ref{partition-pureCS}) below, up to a shift of the level $k$ due to quantum corrections. 
The integral resembles the partition function of a Hermitian matrix model, but the interaction between the eigenvalues is more complicated. 

The matrix model of \cite{Marino:2002fk} was also derived in the context of string theory in \cite{Aganagic:2002wv}. 
In this derivation, pure Chern-Simons theory is realized on D-branes in the topological A-model which wraps $S^3$ in the deformed conifold \cite{Witten:1992fb}. 
By mirror symmetry, this corresponds to D-branes in the topological B-model which wraps $\mathbb{P}^1$ in the mirror geometry. 
According to \cite{Dijkgraaf:2002fc}, the world-volume theory on the D-branes is a matrix model which turns out to coincide with the one of \cite{Marino:2002fk}. 
The planar resolvent of the matrix model was obtained in \cite{Aganagic:2002wv} which reproduces a geometry related the mirror geometry by the large $N$ duality \cite{Gopakumar:1998ki}. 
The free energy of the matrix model is determined explicitly by solving the Picard-Fuchs equations for the geometry obtained from the planar resolvent. 

It is now known that the matrix model of \cite{Marino:2002fk} can be obtained within the context of quantum field theory by using the supersymmetric localization \cite{Kapustin:2009kz}. 
By this method, the vevs of BPS Wilson loops are also given as finite-dimensional integrals, in addition to the partition function. 
Note that there is no shift of the level $k$ in the formulas in \cite{Kapustin:2009kz}. 
It is because the theory considered in \cite{Kapustin:2009kz} is actually ${\cal N}=2$ version of pure Chern-Simons theory in which quantum corrections to the level cancel among them \cite{Kao:1995gf}. 

\vspace{5mm}

\subsection{Planar limit}
   \label{Planar limit}

\vspace{5mm}

Consider ${\cal N}=2$ pure Chern-Simons theory on $S^3$ with the gauge group ${\rm U}(N)$ and the level $k$. 
As reviewed above, the partition function is reduced to 
\begin{equation}
Z\ =\ \int du\,\exp\left( \sum_{i=1}^N\frac{ik}{4\pi}u_i^2 \right)\prod_{i<j}\sinh^2\frac{u_i-u_j}2. 
   \label{partition-pureCS}
\end{equation}
The integral in the right-hand side is a function of $k$ defined on the upper-half plane. 
The partition function is then defined by its boundary value. 

We focus our attention on the planar limit of the theory defined as 
\begin{equation}
N,k\ \to\ \infty, \hspace{1cm} t\ :=\ \frac{2\pi iN}k \hspace{5mm} \mbox{fixed.}
\end{equation}
In the following, we call $t$ the 't~Hooft coupling, instead of $\lambda$ defined above. 
In this limit, the saddle point approximation becomes exact. 
Then, various observables of the theory is determined by an appropriate solution of the saddle point equations 
\begin{equation}
\frac k{2\pi i}u_i\ =\ \sum_{j\ne i}\coth\frac{u_i-u_j}2. 
   \label{saddle-pureCS}
\end{equation}
Let $\{\bar u_i\}$ be a solution. 
The definition (\ref{partition-pureCS}) of the integral shows that $\bar u_i$ will naturally distribute around $u_i=0$ when $k$ is large. 
This is indeed the relevant saddle point which correctly reproduces known results of pure Chern-Simons theory. 
Other solutions of the saddle point equations are discussed in \cite{Morita:2017oev}\cite{Morita:2018oel}. 

We are interested in the vev of the BPS Wilson loop in the fundamental representation. 
In the planar limit, it is given in terms of $\bar u_i$ as 
\begin{equation}
W\ =\ \frac1N\sum_{i=1}^Ne^{\bar u_i}. 
   \label{WL-pureCS-sum}
\end{equation}
We always assume that the planar limit is implicitly taken for expressions like this one. 

\vspace{5mm}

\begin{figure}
\begin{center}
\includegraphics{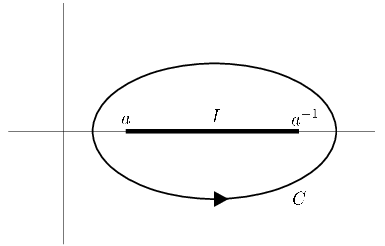}
\end{center}
\caption{
The resolvent $v(z)$ has a branch cut on the line segment $I$ whose endpoints are at $z=a$ and $z=a^{-1}$. 
The integration contour $C$ encircles $I$ counter-clockwise. 
The value $a$ can be complex, so that $I$ does not necessarily lie on the real axis in general. 
}
   \label{fig-pureCS}
\end{figure}

The information of the theory in the planar limit is encoded into the resolvent 
\begin{equation}
v(z)\ :=\ \frac tN\sum_{i=1}^N\frac{z+z_i}{z-z_i}, \hspace{1cm} z_i\ :=\ e^{\bar u_i}. 
   \label{resolvent-pureCS}
\end{equation}
This has a branch cut on a line segment $I$ on which $z_i$ condense. 
By the symmetry of the saddle point equations (\ref{saddle-pureCS}), this satisfies $v(z^{-1})=-v(z)$. 
The endpoints of $I$ are therefore at $z=a,a^{-1}$ for some $a\in\mathbb{C}$. 
We assume $|a|\le1$. 
See Figure \ref{fig-pureCS}. 

The resolvent is determined explicitly in \cite{Aganagic:2002wv}. 
In our notation, it is given as 
\begin{equation}
v(z)\ =\ 2\log\frac{z+1-\sqrt{(z-a)(z-a^{-1})}}{\sqrt{a}+\sqrt{a^{-1}}}. 
\end{equation}
Physical observables can be written in terms of $v(z)$. 
The Wilson loop (\ref{WL-pureCS-sum}) is given as 
\begin{equation}
W\ =\ \frac1{2t}\int_C\frac{dz}{2\pi i}v(z), 
   \label{WL-pureCS-int z}
\end{equation}
where $C$ is a closed contour which encircles the branch cut $I$ counter-clockwise, as depicted in Figure \ref{fig-pureCS}. 
The 't~Hooft coupling $t$ is also given as 
\begin{equation}
t\ =\ \frac12\int_C\frac{dz}{2\pi i}\frac{v(z)}z. 
   \label{tH-v-pureCS}
\end{equation}

\vspace{5mm}

It turns out that the derivative of $v(z)$ has a simpler form \cite{Suyama:2016nap} 
\begin{equation}
zv'(z)-1\ =\ -\frac{z-1}{\sqrt{(z-a)(z-a^{-1})}}. 
\end{equation}
This is a solution of an algebraic equation 
\begin{equation}
\omega^2\ =\ \frac{(z-1)^2}{(z-a)(z-a^{-1})}. 
   \label{curve-pureCS-z}
\end{equation}
Note that this equation can be derived directly from the saddle point equations (\ref{saddle-pureCS}). 
See appendix \ref{curve-derivation}. 
The solution $\omega(z)$ plays the same role as the resolvent $v(z)$. 
For example, the integration by parts in (\ref{WL-pureCS-int z}) gives 
\begin{equation}
W\ =\ -\frac1{2t}\int_C\frac{dz}{2\pi i}\omega(z). 
\end{equation}
In the following, we also consider generalized quantities defined as 
\begin{equation}
W_n\ :=\ \frac1N\sum_{i=1}^Ne^{n\bar u_i}. 
\end{equation}
These can be used to obtain the vev of a BPS Wilson loop in a general representation in the planar limit \cite{Gross:1998gk}. 
These also have integral expressions as 
\begin{equation}
W_n\ =\ -\frac1{2nt}\int_C\frac{dz}{2\pi i}z^{n-1}\omega(z). 
   \label{integral-WL}
\end{equation}
The integration by parts in (\ref{tH-v-pureCS}) gives 
\begin{equation}
t\ =\ -\frac1{4\pi i}\int_C\frac{dz}z\log z\,\omega(z). 
\end{equation}

It turns out that the introduction of a variable 
\begin{equation}
x\ :=\ z+z^{-1} 
\end{equation}
is convenient for the following analysis. 
In terms of $x$, the equation (\ref{curve-pureCS-z}) becomes 
\begin{equation}
y^2\ =\ \frac{x-2}{x-\alpha}, \hspace{1cm} \alpha\ :=\ a+a^{-1}. 
   \label{curve-pureCS}
\end{equation}
The algebraic curve defined by this equation will play an important role in the following. 
The Wilson loop can be written as 
\begin{equation}
W\ =\ -\frac1{4t}\int_{x(C)}\frac{dx}{2\pi i}y(x), 
\end{equation}
where $x(C)$ is the image on the curve of the contour $C$ in the $z$-plane. 
Rewrting $W_n$ is not easy. 
Instead, we consider 
\begin{equation}
\pi_n\ :=\ \int_{x(C)}\frac{dx}{2\pi i}x^ny(x). 
   \label{integral-pi}
\end{equation}
These are linear combinations of $W_n$. 
The expression of $W_n$ in terms of $\pi_n$ will be given later. 

\vspace{5mm}

\subsection{Picard-Fuchs equations}
   \label{PF-pureCS}

\vspace{5mm}

We observed that the Wilson loop $W$ and the 't~Hooft coupling $t$ are given as integrals along a closed contour on an algebraic curve. 
This suggests that they may satisfy some differential equations, which we call Picard-Fuchs equations. 
We will derive these equations for $W$ and $t$ from the curve (\ref{curve-pureCS}), and analyze their solutions. 
A different equation for $t$ obtained in a string theory context is discussed in \cite{Aganagic:2002wv}. 
In the next subsection, we also determine $W_n$ for general $n$ using the properties of the curve (\ref{curve-pureCS}). 

\vspace{5mm}

\subsubsection{Wilson loop}

\vspace{5mm}

First, we derive a differential equation satisfied by 
\begin{equation}
\pi_0\ =\ \int_{x(C)}\frac{dx}{2\pi i}y(x). 
\end{equation}
This is a function of the parameter $\alpha$ defined in (\ref{curve-pureCS}). 
This is related to the Wilson loop $W$ as 
\begin{equation}
\pi_0\ =\ -4tW. 
   \label{pi_0 and W}
\end{equation}
Note that $t$ is also a function of $\alpha$. 

The derivative of $\pi_0$ with respect to $\alpha$ is given by $\partial_\alpha y$. 
The defining equation (\ref{curve-pureCS}) 
of the curve gives a relation 
\begin{equation}
2\partial_\alpha y\ =\ \frac y{x-\alpha}. 
   \label{constraint-pureCS}
\end{equation}
The integration of both sides gives 
\begin{equation}
2\partial_\alpha\pi_0\ =\ \int_{x(C)}\frac{dx}{2\pi i}\frac y{x-\alpha}. 
\end{equation}
In order to obtain a closed equation for $\pi_0$, we need to rewrite the right-hand side in term of $\pi_0$. 

For this purpose, note that the equation (\ref{curve-pureCS}) also gives 
\begin{eqnarray}
2\partial_xy 
&=& \frac y{x-2}-\frac y{x-\alpha}, \\ [2mm] 
2\partial_x(xy) 
&=& 2y+2\frac y{x-2}-\alpha\frac y{x-\alpha}. 
\end{eqnarray}
These relations then give 
\begin{equation}
(\alpha-2)\frac y{x-\alpha}\ =\ 4\partial_xy-2\partial_x(xy)+2y. 
   \label{basis-pureCS}
\end{equation}
Since the integration contour $x(C)$ is closed, the terms $\partial_xy$ and $\partial_x(xy)$ are irrelevant after integration. 
Finally, we obtain 
\begin{equation}
(\alpha-2)\partial_\alpha\pi_0-\pi_0\ =\ 0. 
   \label{PF-pi_0}
\end{equation}

\vspace{5mm}

\subsubsection{'t~Hooft coupling}

\vspace{5mm}

Next, we derive another differential equation satisfied by the 't~Hooft coupling $t$. 
Recall the integral expression 
\begin{equation}
t\ =\ -\frac1{4\pi i}\int_C\frac{dz}z\log z\,y(z+z^{-1}). 
   \label{tH-pureCS}
\end{equation}
From the relation (\ref{constraint-pureCS}), we obtain 
\begin{equation}
2\partial_\alpha t\ =\ -\frac1{4\pi i}\int_C\frac{dz}z\log z\,\frac y{x-\alpha}. 
\end{equation}
We will rewrite the integrand $\log z\,y/(x-\alpha)$ up to terms of the form $z\partial_z(\cdots)$ which do not contribute to $\partial_\alpha t$. 

For this purpose, we start with 
\begin{equation}
2\partial_xy\ =\ \frac y{x-2}-\frac y{x-\alpha}
\end{equation}
which we already used above. 
Recalling $x=z+z^{-1}$, we notice that   
\begin{equation}
2(z-z^{-1})z\partial_zy\ =\ -(\alpha-2)y-(\alpha^2-4)\frac y{x-\alpha} 
\end{equation}
holds. 
We use this to obtain the desired relation 
\begin{eqnarray}
2z\partial_z\Bigl[ (z-z^{-1})\log z\,y \Bigr] 
&=& 2\log z\,xy-(\alpha-2)\log z\,y+2(z-z^{-1})y \nonumber \\ [2mm] 
& & -(\alpha^2-4)\log z\,\frac y{x-\alpha}. 
   \label{constraint1-pureCS}
\end{eqnarray}
The integration of both sides gives 
\begin{equation}
-\frac{\alpha^2-4}{4\pi i}\int_C\frac{dz}z\log z\frac y{x-\alpha}\ =\ 2\tau_1-(\alpha-2)t-\pi_0, 
\end{equation}
where we have defined 
\begin{equation}
\tau_1\ :=\ -\frac1{4\pi i}\int_C\frac{dz}z\log z\,xy(z+z^{-1}). 
\end{equation}
Then, we obtain a differential equation 
\begin{equation}
2(\alpha^2-4)\partial_\alpha t\ =\ 2\tau_1-(\alpha-2)t-\pi_0, 
   \label{PF' tHooft pureCS}
\end{equation}
which is, however, not closed for $t$, even when $\pi_0$ is known. 

Interestingly, the unknown function $\tau_1$ can be obtained from $t$ as follows. 
The relation (\ref{constraint-pureCS}) also gives 
\begin{equation}
2\partial_\alpha(xy)\ =\ y+\alpha\frac y{x-\alpha}. 
\end{equation}
The integration then gives 
\begin{equation}
2\partial_\alpha\tau_1\ =\ t+2\alpha\partial_\alpha t. 
\end{equation}
This can be used to eliminate $\tau_1$ from (\ref{PF' tHooft pureCS}). 
As a result, we obtain 
\begin{equation}
2(\alpha^2-4)\partial_\alpha^2t+(3\alpha-2)\partial_\alpha t\ =\ -\partial_\alpha\pi_0. 
\end{equation}
This is an inhomogeneous equation with the source term given by $\pi_0$. 

\vspace{5mm}

\subsubsection{Solution of the equations}

\vspace{5mm}

Recall that $\pi_0$ satisfies 
\begin{equation}
(\alpha-2)\partial_\alpha\pi_0-\pi_0\ =\ 0. 
\end{equation}
The solution is simply 
\begin{equation}
\pi_0\ =\ A(\alpha-2). 
\end{equation}
The constant $A$ will be determined shortly. 
Then $t$ satisfies 
\begin{equation}
2(\alpha^2-4)\partial_\alpha^2t+(3\alpha-2)\partial_\alpha t\ =\ -\partial_\alpha\pi_0\ =\ -A. 
\end{equation}
This is first-order in $\partial_\alpha t$, so this can be solved easily. 
We obtain 
\begin{equation}
t\ =\ -A\log(\alpha+2)+B\arctan\frac{\sqrt{\alpha-2}}2+\log C. 
\end{equation}

We determine these constants as follows. 
Suppose that $k$ is large, corresponding to small $t$. 
The definition (\ref{partition-pureCS}) of the partition function implies that the saddle point solution $\bar u_i$ become close to zero. 
Then, the branch cut of the resolvent $v(z)$ becomes small, which is realized by the limit $a\to1$ or $\alpha\to2$. 
The relation (\ref{pi_0 and W}) implies that $\pi_0$ should behave as 
\begin{equation}
\pi_0\ \sim\ -4t 
\end{equation}
for small $t$. 
Therefore, we should have $t\sim -A(\alpha-2)/4$ as the leading term. 
This information tells us that $B$ must vanish and $C$ is determined such that $t$ vanishies at $\alpha=2$. 
As a result, we obtain 
\begin{equation}
t\ =\ -A\log\frac{\alpha+2}4. 
\end{equation}

In order to fix the remaining constant $A$, we need an input from the curve (\ref{curve-pureCS}). 
Since the branch cut of $y(z+z^{-1})$ shrinks to a point in the limit $a\to1$, the contour integral (\ref{tH-pureCS}) for $t$ only depends on the local behavior of the curve around $z=1$. 
This is given as 
\begin{equation}
y(z)\ \sim\ -1-\frac 1{2(z-1)^2}(\alpha-2)
\end{equation}
for $\alpha\sim2$. 
This implies that the leading term of $t$ is 
\begin{equation}
t\ \sim\ \frac14(\alpha-2), 
\end{equation}
which then implies $A=-1$. 
In summary, we have found 
\begin{equation}
\pi_0\ =\ 2-\alpha, \hspace{1cm} t\ =\ \log\frac{\alpha+2}4. 
   \label{solution-pureCS}
\end{equation}

As a check, let us compare them with the exact result \cite{Kapustin:2009kz}
\begin{equation}
W\ =\ \frac1Ne^{\pi iN/k}\frac{\sin\pi N/k}{\sin\pi/k}. 
   \label{KWY-WL}
\end{equation}
Note that this is slightly different from (\ref{Witten-WL}) due to the absent of the shift of the level $k$, the inclusion of the framing factor, and the different normalization of the Wilson loop in \cite{Kapustin:2009kz} and \cite{Witten:1988hf}. 
By eliminating $\alpha$, we obtain 
\begin{equation}
W\ =\ \frac{e^t-1}t
\end{equation}
for the Wilson loop . 
This correctly reproduces the planar limit of (\ref{KWY-WL}). 

\vspace{5mm}

\subsection{Recursion relation}
   \label{Recursion relation}

\vspace{5mm}

Let us reconsider the derivation of the Picard-Fuchs equation for $\pi_0$. 
We started with 
\begin{equation}
2\partial_\alpha y\ =\ \frac y{x-\alpha}. 
\end{equation}
We can parametrize $x$ and $y$ near $x=\alpha$ on the curve (\ref{curve-pureCS}) as 
\begin{equation}
x\ \sim\ \alpha+\xi^2, \hspace{1cm} y\ \sim\ \frac{\sqrt{\alpha-2}}\xi. 
\end{equation}
In terms of the local coordinate $\xi$, we find that $\partial_\alpha y$ has a triple pole at $\xi=0$. 
It is holomorphic elsewhere. 
Similarly, 
\begin{equation}
2\partial_x y\ =\ \frac y{x-2}-\frac y{x-\alpha}
\end{equation}
shows that $\partial_xy$ also has a triple pole at $x=\alpha$. 
The parametrization around $x=2$,  
\begin{equation}
x\ \sim\ 2+\eta^2, \hspace{1cm} y\ \sim\ \frac \eta{\sqrt{2-\alpha}}, 
\end{equation}
implies that $\partial_xy$ has a simple pole at $\eta=0$. 
Otherwise it is holomorphic. 

Let $V$ be a vector space of meromorphic functions on the curve (\ref{curve-pureCS}), each member of which has at most a triple pole at $x=\alpha$ and at most a simple pole at $x=2$, and is holomorphic elsewhere. 
It is known that the dimension of $V$ is finite. 
The multiplication by $x$ does not increase the orders of poles at $x=2,\alpha$. 
This operation may produce poles at $x=\infty$. 
After subtracting the poles at infinity, the resulting function is again a member of $V$. 
For example, 
\begin{equation}
2x\partial_xy\ =\ 2\frac y{x-2}-\alpha\frac y{x-\alpha}\ \in\ V. 
\end{equation}
If there are more than $\dim V$ functions in $V$, then they are linearly dependent. 

Our situation is much simpler. 
We find that three functions $\partial_\alpha y$, $\partial_xy$, $\partial_x(xy)-y$ form a two-dimensional subspace $V_2$ of $V$ for which 
\begin{equation}
\frac y{x-2}, \hspace{1cm} \frac y{x-\alpha}
\end{equation}
is a basis. 
Therefore, there must exist a non-trivial relation among them, namely 
\begin{equation}
(\alpha-2)\partial_\alpha y\ =\ 2\partial_xy-\Bigl( \partial_x(xy)-y \Bigr), 
\end{equation}
from which we can derive the Picard-Fuchs equation (\ref{PF-pi_0}). 

\vspace{5mm}

In a similar manner, we can obtain an infinite number of non-trivial relations. 
For example, we find that 
\begin{equation}
2\partial_x(x^ny)\ =\ S_n+2^n\frac y{x-2}-\alpha^n\frac y{x-\alpha}
   \label{partial_x(x^ny)}
\end{equation}
holds, where 
\begin{eqnarray}
S_n 
& :=& 2nx^{n-1}y+\frac{x^n-2^n}{x-2}y-\frac{x^n-\alpha^n}{x-\alpha}y \nonumber \\
&=& 2nx^{n-1}y+\sum_{k=0}^{n-1}(2^{n-1-k}-\alpha^{n-1-k})x^ky. 
\end{eqnarray}
All of the functions $\partial_x(x^ny)-S_n$ therefore belong to $V_2$, so any three of them are linearly dependent. 
We can use these relations to obtain a recursion relation which determines $\pi_n$ for general $n$ as follows. 

The relation (\ref{partial_x(x^ny)}) can be organized in the form 
\begin{equation}
\left[
\begin{array}{c}
\partial_x(x^ny)-S_n \\ [2mm] 
\partial_x(x^{n+1}y)-S_{n+1}
\end{array}
\right] 
\ =\ 
\left[
\begin{array}{cc}
2^n & \alpha^n \\ [2mm] 
2^{n+1} & \alpha^{n+1}
\end{array}
\right]
\left[
\begin{array}{c}
\frac y{x-2} \\ [2mm] 
\frac {-y}{x-\alpha}
\end{array}
\right]. 
\end{equation}
Using an identity 
\begin{equation}
(-2\alpha,\alpha+2)
\left[
\begin{array}{cc}
2^n & \alpha^n \\
2^{n+1} & \alpha^{n+1}
\end{array}
\right]
\ =\ (2^{n+2},\alpha^{n+2}), 
\end{equation}
we find 
\begin{equation}
2\alpha S_n-(\alpha+2)S_{n+1}+S_{n+2}\ =\ \partial_x(\cdots). 
\end{equation}
The integration of both sides gives a three-term recursion relation 
\begin{equation}
4\alpha n\,\pi_{n-1}-\Bigl( 2(\alpha+2)(n+1)+\alpha-2 \Bigr)\pi_n+2(n+2)\pi_{n+1}\ =\ 0. 
   \label{recursion-pureCS}
\end{equation}

We need to determine $\pi_1$ for this recursion relation to work. 
Note that the above calculation with $n=0$ gives 
\begin{equation}
-(\alpha+2)S_1+S_2\ =\ -2(\alpha+2)y+4xy+(2-\alpha)y\ =\ \partial_x(\cdots). 
\end{equation}
This implies 
\begin{equation}
\pi_1\ =\ \frac14(3\alpha+2)\pi_0\ =\ \frac14(3\alpha+2)(2-\alpha). 
\end{equation}
Then, the recursion relation gives 
\begin{equation}
\pi_2\ =\ -\frac16\left( 4\alpha\pi_0-(5\alpha+6)\pi_1 \right)\ =\ -\frac18(\alpha-2)(5\alpha^2+4\alpha+4) 
\end{equation}
and so on. 

Remarkably, the generating function of $\pi_n$ defined as 
\begin{equation}
g(t)\ :=\ \sum_{n=0}^\infty \pi_nt^{n+1}
   \label{generating fn-pureCS}
\end{equation}
can be determined explicitly. 
The recursion relation (\ref{recursion-pureCS}) implies that $g(t)$ satisfies 
\begin{equation}
2(2t-1)(\alpha t-1)\partial_tg-(\alpha-2)g\ =\ \Bigl( -(3\alpha+2)t+2 \Bigr)\pi_0+4t\pi_1. 
\end{equation}
The solution satisfying $g(0)=0$ is simply 
\begin{equation}
g(t)\ =\ 2-2\sqrt{\frac{1-2t}{1-\alpha t}}. 
\end{equation}

This function is actually related to $y(x)$. 
By the definition of $g(t)$, we have 
\begin{equation}
g(t)\ =\ \int_{x(C)}\frac{dx}{2\pi i}\frac t{1-tx}y(x)\ =\ 2+2y(t^{-1}). 
\end{equation}
This gives the expressions of $W_n$ in terms of $\pi_n$. 
Recall that $y(x)$ is obtained from the resolvent $v(z)$ defined as in (\ref{resolvent-pureCS}) which has the expansion 
\begin{equation}
v(z)\ =\ t+2\sum_{n=1}^\infty tW_nz^{-n} 
   \label{v(z) expansion}
\end{equation}
given in terms of $W_n$. 
Then, $y(x)$ is given as 
\begin{equation}
2+2y(x)\ =\ -4\sum_{n=1}^\infty ntW_nz^{-n}. 
\end{equation}
By comparing this expansion with 
\begin{equation}
g(x^{-1})\ =\ \sum_{n=0}^\infty \pi_n\left( z+z^{-1} \right)^{n+1}, 
\end{equation}
we obtain the desired relations. 
For example, we obtain 
\begin{equation}
W_1\ =\ -\frac1{4t}\pi_0, \hspace{1cm} W_2\ =\ -\frac1{8t}\pi_1, \hspace{1cm} W_3\ =\ -\frac1{12t}(\pi_2-\pi_0),  \end{equation}
and so on. 
Note that the integral expressions (\ref{integral-WL}) and (\ref{integral-pi}) for $W_n$ and $\pi_n$, respectively, are formally independent of the theory under consideration. 
Therefore, the same relation between $W_n$ and $\pi_n$ can be also used for ABJM theory. 

\vspace{5mm}

In fact, the information on $W_n$ can be obtained more easily and more explicitly for pure Chern-Simons theory. 
Recall 
\begin{equation}
zv'(z)-1\ =\ -\frac{z-1}{\sqrt{(z-a)(z-a^{-1})}}. 
\end{equation}
Using the expansion (\ref{v(z) expansion}), we find 
\begin{equation}
tW_n\ =\ \frac1{2n}\left[ P_n\left( \frac\alpha2 \right)-P_{n-1}\left( \frac\alpha2 \right) \right], 
\end{equation}
where $P_n(x)$ are Legendre polynomials. 

In the next section, we investigate ABJM theory for which the Picard-Fuchs equations may give us results which is not easily deduced from the resolvent. 

\vspace{1cm}

\section{ABJM theory}
   \label{ABJM theory}

\vspace{5mm}

In this section, we extend the investigation of Picard-Fuchs equations to ABJM theory. 
As a starting point, we consider a more general theory including ABJ theory and GT theory \cite{Gaiotto:2009mv}. 
The gauge group of the theory is ${\rm U}(N_1)\times{\rm U}(N_2)$, and the levels are $k_1$ and $k_2$. 
ABJM theory corresponds to $N_1=N_2$ and $k_1=-k_2$. 

The partition function is reduced to \cite{Kapustin:2009kz} 
\begin{equation}
Z\ =\ \int dud\tilde u\,\exp\left[ \sum_{i=1}^{N_1}\frac{ik_1}{4\pi}(u_{1i})^2+\sum_{j=1}^{N_2}\frac{ik_2}{4\pi}(u_{2j})^2 \right]\frac{\displaystyle{\prod_{i<i'}{\sinh^2\frac{u_{1i}-u_{1i'}}2}\prod_{j<j'}\sinh^2\frac{u_{2j}-u_{2j'}}2}}{\displaystyle\prod_{i=1}^{N_1}\prod_{j=1}^{N_2}\cosh^2\frac{u_{1i}-u_{2j}}2}. 
   \label{partition-ABJM}
\end{equation}
As for pure Chern-Simons theory, this integral is a function of $k_1$ and $k_2$ which is defined when both $k_1$ and $k_2$ are in the upper-half plane. 

In order to define the planar limit, we introduce an auxiliary parameter $k$ which is taken to be large. 
The parameters $N_1,N_2,k_1,k_2$ are assumed to be proportional to $k$. 
In other words, we take the large $k$ limit while keeping the ratios 
\begin{equation}
t_1\ :=\ \frac{2\pi iN_1}k, \hspace{1cm} t_2\ :=\ \frac{2\pi iN_2}k, \hspace{1cm} \kappa_1\ :=\ \frac{k_1}k, \hspace{1cm} \kappa_2\ :=\ \frac{k_2}k
\end{equation}
fixed. 
For ABJM theory, we have $t_1=t_2$. 
In the following, we choose $k$ such that $\kappa_1=-\kappa_2=1$ holds. 

We are interested in the vevs of BPS Wilson loops. 
As reviewed in the introduction, there are bosonic BPS Wilson loops \cite{Drukker:2008zx}\cite{Chen:2008bp}\cite{Rey:2008bh} in the fundamental representation of ${\rm U}(N_1)$ or ${\rm U}(N_2)$. 
Their vevs are given in terms of the saddle point $\{\bar u_{1i},\bar u_{2j}\}$ of the integral (\ref{partition-ABJM}) as 
\begin{equation}
W_1\ =\ \frac1{N_1}\sum_{i=1}^{N_1}e^{\bar u_{1i}}, \hspace{1cm} W_2\ =\ \frac1{N_2}\sum_{j=1}^{N_2}e^{\bar u_{2j}}. 
\end{equation}
There is also a fermionic BPS Wilson loop \cite{Drukker:2009hy}. 
The vev turns out to be given as $N_1W_1+N_2W_2$. 
In the planar limit, it is more appropriate to consider 
\begin{equation}
{\cal W}\ =\ t_1W_1+t_2W_2. 
\end{equation}

\begin{figure}
\begin{center}
\includegraphics{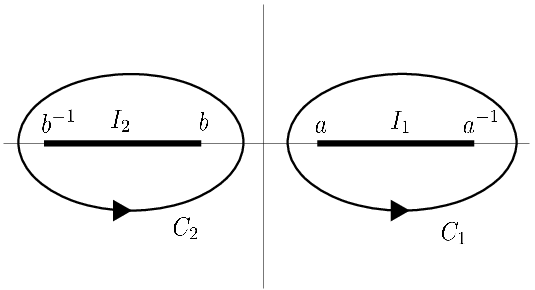}
\end{center}
\caption{
The resolvent $v(z)$ has two branch cuts as shown in the figure. 
The integration contour $C_1$ encircles the cut $I_1$ counter-clockwise, and $C_2$ encircles $I_2$ similarly. 
The cuts do not necessarily lie on the real axis. 
}
   \label{fig-ABJM}
\end{figure}

In order to analyze these observables, we define the resolvents 
\begin{eqnarray}
v_1(z) 
&:=& \frac{t_1}{N_1}\sum_{i=1}^{N_1}\frac{z+z_{1i}}{z-z_{1i}}, \hspace{1cm} z_{1i}\ :=\ e^{\bar u_{1i}}, 
   \label{resolvent1-ABJM} \\
v_2(z) 
&:=& \frac{t_2}{N_2}\sum_{j=1}^{N_2}\frac{z+z_{2j}}{z-z_{2j}}, \hspace{1cm} z_{2j}\ :=\ -e^{\bar u_{2j}}.  
   \label{resolvent2-ABJM}
\end{eqnarray}
Interestingly, a combination $v(z):=v_1(z)-v_2(z)$ of them can be determined explicitly as \cite{Halmagyi:2003ze}\cite{Marino:2009jd} 
\begin{equation}
v(z)\ =\ 2\log\left[ \frac{\sqrt{(z-b)(z-b^{-1})}-\sqrt{(z-a)(z-a^{-1})}}{\sqrt{a+a^{-1}-b-b^{-1}}} \right]. 
\end{equation}
This satisfies $v(z^{-1})=-v(z)$. 
The function $v(z)$ has two branch cuts on the line segments $I_1$ and $I_2$ on which $z_{1i}$ and $z_{2j}$, respectively, condense. 
See Figure \ref{fig-ABJM}. 
We assume $|a|,|b|\le1$. 

Instead of $v(z)$, it is easier to analyze its derivative 
\begin{equation}
zv'(z)-1\ =\ -\frac{z^2-1}{\sqrt{(z-a)(z-a^{-1})(z-b)(z-b^{-1})}}. 
\end{equation}
As for pure Chern-Simons theory, we introduce the variable $x=z+z^{-1}$ and consider an algebraic curve defined by 
\begin{equation}
y^2\ =\ \frac{(x-2)(x+2)}{(x-\alpha)(x-\beta)}, 
   \label{curve-ABJM}
\end{equation}
where $\alpha=a+a^{-1}$ and $\beta=b+b^{-1}$. 
The derivative $zv'(z)$ is then given in terms of $y(x)$ as 
\begin{equation}
zv'(z)\ =\ 1+y(z+z^{-1}). 
\end{equation}
A derivation of the curve from the saddle point of (\ref{partition-ABJM}) is summarized in appendix \ref{curve-derivation}. 

The vevs of the BPS Wilson loops have integral representations 
\begin{equation}
W_1\ =\ -\frac1{4t_1}\int_{C_1}\frac{dx}{2\pi i}y(x), \hspace{1cm} W_2\ =\ -\frac1{4t_2}\int_{C_2}\frac{dx}{2\pi i}y(x), 
\end{equation}
and 
\begin{equation}
{\cal W}\ =\ -\frac14\int_{C_1+C_2}\frac{dx}{2\pi i}y(x). 
\end{equation}
The integration contours $C_1$ and $C_2$ are depicted in Figure \ref{fig-ABJM}. 
Note that these Wilson loops are given by essentially the same integral, and the different choice of the integration contour gives the different Wilson loop. 

The integral representations for the 't~Hooft couplings 
\begin{equation}
t_1\ =\ -\frac1{4\pi i}\int_{C_1}\frac{dz}z\log z\,y(z+z^{-1}), \hspace{1cm} t_2\ =\ \frac1{4\pi i}\int_{C_2}\frac{dz}z\log z\,y(z+z^{-1})  
\end{equation}
can be used to relate the parameters $\alpha, \beta$ to $t_1, t_2$. 

\vspace{5mm}

In the following, we focus our attention to ABJM theory. 
Accordingly, we will restrict the parameter space $\mathbb{C}^2$ to a subspace, usually called the ABJM slice. 
This corresponds to $t_1=t_2$. 
It turns out that $t_1-t_2$ can be given simply \cite{Marino:2009jd} 
\begin{equation}
t_1-t_2\ =\ \frac12\int_0^\infty\frac{dz}z\Bigl( y(z+z^{-1})+1 \Bigr)\ =\ \log\frac{\alpha-\beta}4. 
\end{equation}
This implies that the ABJM slice is parametrized by $\gamma\in\mathbb{C}$ as 
\begin{equation}
\alpha\ =\ 2+\gamma, \hspace{1cm} \beta\ =\ -2+\gamma. 
\end{equation}
In the following, we denote $t=t_1=t_2$ for simplicity. 

\vspace{5mm}

\subsection{Picard-Fuchs equations}
   \label{PF-ABJM}

\vspace{5mm}

We derive the Picard-Fuchs equations for the Wilson loops and the 't~Hooft coupling. 
As for pure Chern-Simons theory, we consider instead 
\begin{equation}
\pi_n\ =\int_{x(C)}\frac{dx}{2\pi i}x^ny(x). 
   \label{def-pi_n}
\end{equation}
It turns out that $\pi_n$ satisfy a system of differential equations in which $\pi_n$ couple among them. 
The equations are independent of the choice of the integration contour $C$. 
As a result, the vevs of the bosonic and fermionic Wilson loops are solutions of the same equation. 
A geometric description of $W_n$ which are linear combinations of $\pi_n$ is given in \cite{Klemm:2012ii}. 

\vspace{5mm}

\subsubsection{A basis of functions}

\vspace{5mm}

From (\ref{curve-ABJM}), we obtain 
\begin{equation}
2\partial_xy\ =\ \frac y{x-2}+\frac y{x+2}-\frac y{x-\alpha}-\frac y{x-\beta}. 
   \label{constraint1-ABJM}
\end{equation}
Based on the experience in pure Chern-Simons theory, we should consider a four-dimensional subspace $V_4$, spanned by 
\begin{equation}
\frac y{x-2}, \hspace{5mm} \frac y{x+2}, \hspace{5mm} \frac {-y}{x-\alpha}, \hspace{5mm} \frac {-y}{x-\beta}, 
   \label{basis fns-ABJM}
\end{equation}
of the vector space of meromorphic functions on the curve (\ref{curve-ABJM}). 
We would like to express these basis functions as linear combinations of terms of the form $x^ny$, up to total derivatives, like the expression (\ref{basis-pureCS}) for pure Chern-Simons theory. 

For this purpose, we use 
\begin{eqnarray}
2\partial_x(x^ny) 
&=& 2nx^{n-1}y+\frac{x^n-2^n}{x-2}y+\frac{x^n-(-2)^n}{x+2}y-\frac{x^n-\alpha^n}{x-\alpha}y-\frac{x^n-\beta^n}{x-\beta}y \nonumber \\ [2mm] 
& & +2^n\frac y{x-2}+(-2)^n\frac y{x+2}-\alpha^n\frac y{x-\alpha}-\beta^n\frac y{x-\beta}, 
\end{eqnarray}
which is derived from (\ref{constraint1-ABJM}). 
Note that the first line in the right-hand side is 
\begin{eqnarray}
& & \frac{x^n-2^n}{x-2}y+\frac{x^n-(-2)^n}{x+2}y-\frac{x^n-\alpha^n}{x-\alpha}y-\frac{x^n-\beta^n}{x-\beta}y \nonumber \\
&=& \sum_{k=0}^{n-1}\left( 2^{n-1-k}+(-2)^{n-1-k}-\alpha^{n-1-k}-\beta^{n-1-k} \right)x^ky. 
\end{eqnarray}
We use four of these relations corresponding to $n=0,1,2,3$ to rewrite the basis functions (\ref{basis fns-ABJM}). 
In order to organize these equations, we introduce vectors 
\begin{equation}
Y\ :=\ \left[
\begin{array}{c}
y \\
xy \\
x^2y \\
x^3y
\end{array}
\right], \hspace{1cm} 
B\ :=\ 
\left[
\begin{array}{c}
\frac y{x-2} \\
\frac y{x+2} \\
\frac{-y}{x-\alpha} \\
\frac{-y}{x-\beta}
\end{array}
\right]. 
\end{equation}
We find that they satisfy 
\begin{equation}
2\partial_xY\ =\ TY+UB, 
\end{equation}
where 
\begin{equation}
T\ :=\ 
\left[
\begin{array}{cccc}
0 & 0 & 0 & 0 \\
2 & 0 & 0 & 0 \\
-\alpha-\beta & 4 & 0 & 0 \\
8-\alpha^2-\beta^2 & -\alpha-\beta & 6 & 0
\end{array}
\right], \hspace{1cm} 
U\ :=\ 
\left[
\begin{array}{cccc}
1 & 1 & 1 & 1 \\
2 & -2 & \alpha & \beta \\
2^2 & (-2)^2 & \alpha^2 & \beta^2 \\
2^3 & (-2)^3 & \alpha^3 & \beta^3
\end{array}
\right]. 
\end{equation}

Then, the basis functions are given as 
\begin{equation}
B\ =\ -U^{-1}TY+2U^{-1}\partial_xY. 
   \label{basis-ABJM}
\end{equation}
Note that $U$ is invertible if and only if the Vandermonde determinant $\det U$ is non-zero. 
When $\det U=0$, the dimension of the vector space $V_4$ decreases, and the curve (\ref{curve-ABJM}) degenerates. 
We will see that this degeneration corresponds to singularities of the resulting Picard-Fuchs equations. 

\vspace{5mm}

\subsubsection{Wilson loops}

\vspace{5mm}

We restrict our analysis on the ABJM slice 
\begin{equation}
\alpha\ =\ 2+\gamma, \hspace{1cm} \beta\ =\ -2+\gamma. 
\end{equation}
The derivative of the Wilson loop $W$, or $\pi_0$, with respect to $\gamma$ is given by 
\begin{equation}
2\partial_\gamma y\ =\ \frac y{x-\alpha}+\frac y{x-\beta}. 
\end{equation}
The relation (\ref{basis-ABJM}) implies that $\pi_0$ couples to other $\pi_n$. 
Therefore, we also need to consider $\partial_\gamma\pi_n$, or $\partial_\gamma(x^ny)$ which are given as 
\begin{eqnarray}
2\partial_\gamma(x^ny) 
&=& \frac{x^n-\alpha^n}{x-\alpha}y+\frac{x^n-\beta^n}{x-\beta}y +\alpha^n\frac y{x-\alpha}y+\beta^n\frac y{x-\beta}y. 
\end{eqnarray}
Four of them corresponding to $n=0,1,2,3$ can be organized as 
\begin{equation}
2\partial_\gamma Y\ =\ \widetilde TY+\widetilde UB, 
\end{equation}
where 
\begin{equation}
\widetilde T\ :=\ 
\left[
\begin{array}{cccc}
0 & 0 & 0 & 0 \\
2 & 0 & 0 & 0 \\
\alpha+\beta & 2 & 0 & 0 \\
\alpha^2+\beta^2 & \alpha+\beta & 2 & 0
\end{array}
\right], 
\hspace{1cm} 
\widetilde U\ :=\ 
\left[
\begin{array}{cccc}
0 & 0 & -1 & -1 \\
0 & 0 & -\alpha & -\beta \\
0 & 0 & -\alpha^2 & -\beta^2 \\
0 & 0 & -\alpha^3 & -\beta^3
\end{array}
\right]. 
\end{equation}

Combining this equation with (\ref{basis-ABJM}), we obtain 
\begin{eqnarray}
2\partial_\gamma Y 
&=& \widetilde T Y-\widetilde UU^{-1}TY+2\widetilde UU^{-1}\partial_xY. 
\end{eqnarray}
The integration of both sides gives a system of differential equations for $\pi_0,\cdots,\pi_3$. 
The last term in the right-hand side is irrelevant. 
We find 
\begin{equation}
\widetilde T-\widetilde UU^{-1}T\ =\ \frac2{\gamma(\gamma-4)(\gamma+4)}
\left[
\begin{array}{cccc}
8-\gamma^2 & 8\gamma & -6 & 0 \\
-4\gamma & -16+5\gamma^2 & -3\gamma & 0 \\
32-4\gamma^2 & -16\gamma+3\gamma^3 & -24 & 0 \\
48\gamma-4\gamma^3 & -64+20\gamma^2 & -76\gamma+4\gamma^3 & 0
\end{array}
\right]. 
\end{equation}
This shows that $\pi_0,\pi_1,\pi_2$ decouple from $\pi_3$. 

After a slight modification, we obtain 
\begin{equation}
\partial_\gamma\Pi\ =\ \frac1{\gamma(\gamma-4)(\gamma+4)}{\cal A}\,\Pi, 
   \label{Fuchsian}
\end{equation}
where 
\begin{equation}
\Pi\ :=\ 
\left[
\begin{array}{c}
\pi_0 \\
\pi_1 \\
\pi_2-\gamma\pi_1
\end{array}
\right]
, \hspace{1cm} {\cal A}\ :=\ 
\left[
\begin{array}{cccc}
8-\gamma^2 & 2\gamma & -6 \\
-4\gamma & -16+2\gamma^2 & -3\gamma \\
32 & -8\gamma & -24+3\gamma^2
\end{array}
\right]
\end{equation}
This is a Fuchsian system of differential equations with the regular singularities at $\gamma=0,\pm4,\infty$. 
The remaining \(\pi_3\) is then given in terms of $\pi_0,\pi_1,\pi_2$ as 
\begin{equation}
\partial_\gamma\pi_3\ =\ \frac1{\gamma(\gamma-4)(\gamma+4)}\Bigl[ (48\gamma-4\gamma^3)\pi_0+(-64+20\gamma^2)\pi_1+(-76\gamma+4\gamma^3)\pi_2 \Bigr]. 
\end{equation}

Note that the regular singularities correspond to 
\begin{equation}
\begin{array}{lcll}
\gamma\ =\ 0 & \hspace{5mm} \Leftrightarrow \hspace{5mm} & \alpha\ =\ 2, & \beta\ =\ -2, \\ [1mm] 
\gamma\ =\ 4 & \Leftrightarrow & \alpha\ =\ 6, & \beta\ =\ 2, \\ [1mm] 
\gamma\ =\ -4 & \Leftrightarrow & \alpha\ =\ -2, & \beta\ =\ -6, \\ [1mm] 
\gamma\ =\ \infty & \Leftrightarrow & \alpha\ =\ \infty, & \beta\ =\ \infty. 
\end{array}
\end{equation}
They correspond to various degenerations of the curve (\ref{curve-ABJM}). 
These points on the ABJM slice are discussed in \cite{Marino:2009jd}\cite{Drukker:2010nc}. 
For example, $\gamma=0$ corresponds to the weak coupling limit, while $\gamma=\infty$ describes the strong coupling limit or the gravity dual. 

\vspace{5mm}

The Fuchsian system (\ref{Fuchsian}) can be reduced to a third-order differential equation for $\pi_0$ only, which is easier to deal with. 
As a result, we obtain 
\begin{equation}
\gamma(\gamma-4)(\gamma+4)\partial_\gamma^3\pi_0+2\gamma^2\partial_\gamma^2\pi_0-\gamma\partial_\gamma\pi_0+\pi_0\ =\ 0. 
   \label{3rd-order-WL}
\end{equation}
The solution of this equation then gives $\pi_1$ and $\pi_2$ as 
\begin{eqnarray}
\pi_1 
&=& -\frac12\left[ \gamma(\gamma-4)(\gamma+4)\partial_\gamma^2\pi_0-2\gamma\pi_0 \right], \\ 
\pi_2 
&=& \gamma\pi_1-\frac16\left[ \gamma^2(\gamma-4)(\gamma+4)\partial_\gamma^2\pi_0+\gamma(\gamma-4)(\gamma+4)\partial_\gamma\pi_0-(\gamma^2+8)\pi_0 \right]. 
\end{eqnarray}

We notice that $\pi_0=\gamma$ is a solution of (\ref{3rd-order-WL}). 
This solution corresponds to the vev $\cal W$ of the fermionic BPS Wilson loop which is proportional to $\alpha+\beta$ \cite{Marino:2009jd}. 
This is actually expected since, as was mentioned in the beginning of this subsection, the vevs of both the bosonic and the fermionic Wilson loops are solutions of the same equation (\ref{3rd-order-WL}). 

Using this simple solution, we can further reduce the equation (\ref{3rd-order-WL}). 
Let $\pi_0=\gamma\tilde\pi_0$. 
Then, we find that $\pi:=\partial_\gamma\tilde\pi_0$ satisfies 
\begin{equation}
\gamma(\gamma-4)(\gamma+4)\partial_\gamma^2\pi+(5\gamma^2-48)\partial_\gamma\pi+3\gamma\pi\ =\ 0. 
   \label{2nd-order-WL}
\end{equation}
This is a Heun equation. 
For a review, see \cite{Heun} for example. 
Indeed, by introducing new variables $\gamma=4z$ and $\pi=16y$, we obtain the standard form of the Heun equation 
\begin{equation}
\partial_z^2y+\left[ \frac3z+\frac1{z-1}+\frac1{z+1} \right]\partial_zy+\frac{3z}{z(z-1)(z+1)}y\ =\ 0. 
\end{equation}
Let $f$ and $g$ be a basis of the space of solutions of (\ref{2nd-order-WL}). 
Then, $\pi_0$ is given by their integrals as 
\begin{equation}
\pi_0\ =\ A\gamma+B\gamma\int d\gamma\,f(\gamma)+C\gamma\int d\gamma\,g(\gamma) 
\end{equation}
with appropriate constants $A,B,C$. 

\vspace{5mm}

\subsubsection{'t~Hooft coupling}

\vspace{5mm}

Next, we derive the Picard-Fuchs equation for the 't~Hooft coupling $t$. 
As for $\pi_n$, we need to introduce auxiliary quantities 
\begin{equation}
\tau_n\ :=\ -\frac1{4\pi i}\int_C\frac{dz}z\log z\,x^ny(x), \hspace{1cm} \tau_0\ =\ t, 
\end{equation}
in order to obtain a closed system of equations. 

The derivation is similar to the one for pure Chern-Simons theory. 
The counterpart of (\ref{constraint1-pureCS}) for ABJM theory turns out to be 
\begin{eqnarray}
2z\partial_z\left[ (z-z^{-1})\log z\,y \right] 
&=& 2\log z\,xy-(\alpha+\beta)\log z\,y+2(z-z^{-1})y \nonumber \\
& & -(\alpha^2-4)\frac{\log z}{x-\alpha}y-(\beta^2-4)\frac{\log z}{x-\beta}y, \\ [2mm] 
2z\partial_z\left[ (z-z^{-1})\log z\,xy \right] 
&=& 4\log z\,x^2y-(\alpha+\beta)\log z\,xy-(\alpha^2+\beta^2)\log z\,y+2(z-z^{-1})xy \nonumber \\
& & -\alpha(\alpha^2-4)\frac{\log z}{x-\alpha}y-\beta(\beta^2-4)\frac{\log z}{x-\beta}y. 
\end{eqnarray}
Note that terms proportional to $y/(x\mp2)$ do not appear in the right-hand side of the above relations, although they appear in $\partial_xy$. 
Therefore, these two equations are enough to rewrite the functions 
\begin{equation}
\frac{\log z}{x-\alpha}y, \hspace{1cm} \frac{\log z}{x-\beta}
   \label{basis-tH-ABJM}
\end{equation}
by terms whose integration give $\tau_n$ and $\pi_n$. 
Since there appear terms $\log z\,x^ny$ with $n=0,1,2$, we need to consider the following three $\gamma$-derivatives 
\begin{eqnarray}
2\partial_\gamma(\log z\,y) 
&=& \frac{\log z}{x-\alpha}y+\frac{\log z}{x-\beta}y, \\
2\partial_\gamma(\log z\,xy) 
&=& 2\log z\,y+\alpha\frac{\log z}{x-\alpha}y+\beta\frac{\log z}{x-\beta}y, \\
2\partial_\gamma(\log z\,x^2y) 
&=& 2\log z\,xy+(\alpha+\beta)\log z\,y+\alpha^2\frac{\log z}{x-\alpha}y+\beta^2\frac{\log z}{x-\beta}y, 
\end{eqnarray}
for obtaining a closed set of equations. 
Eliminating the functions (\ref{basis-tH-ABJM}) from the right-hand side and integrating both sides, we obtain a system of differential equations for $\tau_0=t,\tau_1,\tau_2$. 

As a result, we find 
\begin{eqnarray}
\partial_\gamma
\left[
\begin{array}{c}
\tau_0 \\
\tau_1 \\
\tau_2
\end{array}
\right]
&=& \frac1{\gamma(\gamma-4)(\gamma+4)}
\left[
\begin{array}{ccc}
8-\gamma^2 & 5\gamma & -4 \\
-4\gamma & -8+3\gamma^2 & -2\gamma \\
32-4\gamma^2 & -12\gamma+2\gamma^3 & -16
\end{array}
\right]
\left[
\begin{array}{c}
\tau_0 \\
\tau_1 \\
\tau_2
\end{array}
\right] 
\nonumber \\
& & +\frac1{2\gamma(\gamma-4)(\gamma+4)}
\left[
\begin{array}{cc}
-3\gamma & 2 \\
8-2\gamma^2 & \gamma \\
4\gamma-\gamma^3 & 8
\end{array}
\right]
\left[
\begin{array}{c}
\pi_0 \\
\pi_1
\end{array}
\right]. 
\end{eqnarray}
These equations can be reduced to 
\begin{eqnarray}
& & 4\gamma^2(\gamma-4)(\gamma+4)\partial_\gamma^3\tau_0+16\gamma(\gamma-2)(\gamma+2)\partial_\gamma^2\tau_0+8(\gamma^2+2)\partial_\gamma\tau_0 \nonumber \\ [2mm] 
&=& -(3\gamma^2+16)\partial_\gamma^2\pi_0-\gamma\partial_\gamma\pi_0+\pi_0. 
\end{eqnarray}
This is an inhomogeneous Heun equation for $\partial_\gamma\tau_0$, and the source term is given by $\pi_0$. 
Note that the right-hand side can be simplified as 
\begin{equation}
-\gamma(3\gamma^2+16)\partial_\gamma\pi-(7\gamma^2+32)\pi. 
\end{equation}

For completeness, we give the formulas for $\tau_1$ and $\tau_2$ as 
\begin{eqnarray}
\tau_1 
&=& -2\gamma(\gamma-4)(\gamma+4)\partial_\gamma^2\tau_0-(3\gamma^2-16)\partial_\gamma\tau_0+\gamma\tau_0 \nonumber \\
& & -(\gamma-4)(\gamma+4)\partial_\gamma^2\pi_0-2\gamma\partial_\gamma\pi_0+\frac32\pi_0, \\ [2mm] 
\tau_2 
&=& -\frac52\gamma^2(\gamma-4)(\gamma+4)\partial_\gamma^2\tau_0-4\gamma(\gamma^2-6)\partial_\gamma\tau_0+(\gamma^2+2)\tau_0 \nonumber \\
& & -\frac{11}8\gamma(\gamma-4)(\gamma+4)\partial_\gamma^2\pi_0-\frac52\gamma^2\partial_\gamma\pi_0+\frac74\gamma\pi_0. 
\end{eqnarray}
The physical meaning of these quantities are not clear. 

\vspace{5mm}

\subsubsection{Local solutions of the equations}

\vspace{5mm}

We investigate the local solutions of the equations 
\begin{equation}
\gamma(\gamma-4)(\gamma+4)\partial_\gamma^2\pi+(5\gamma^2-48)\partial_\gamma\pi+3\gamma\pi\ =\ 0 
   \label{PF4WL-local}
\end{equation}
and 
\begin{eqnarray}
& & 4\gamma^2(\gamma-4)(\gamma+4)\partial_\gamma^3t+16\gamma(\gamma-2)(\gamma+2)\partial_\gamma^2t+8(\gamma^2+2)\partial_\gamma t \nonumber \\ [2mm] 
&=& -\gamma(3\gamma^2+16)\partial_\gamma\pi-(7\gamma^2+32)\pi 
   \label{PF4tH-local}
\end{eqnarray}
around the regular singularities $\gamma=0$ and $\gamma=\infty$, corresponding to the perturbative limit and the dual gravity limit, respectively. 
The other regular singularities $\gamma=\pm4$ correspond to the ones investigated in \cite{Drukker:2010nc}. 
They are related to pure Chern-Simons theory, which can be understood by comparing the curves (\ref{curve-pureCS}) and (\ref{curve-ABJM}). 

Recall that $\pi_0$ is given in terms of a basis of solutions $f$ and $g$ of (\ref{PF4WL-local}) as 
\begin{equation}
\pi_0\ =\ A\gamma+\gamma\int d\gamma\,\pi(\gamma), \hspace{1cm} \pi\ =\ Bf+Cg. 
\end{equation}

\vspace{5mm}

(i) $\gamma=0$

\vspace{5mm}

First, we consider the local solution of (\ref{PF4WL-local}) around $\gamma=0$. 
The characteristic exponents at $\gamma=0$ are $0$ and $-2$. 
The local solution for the exponent $0$ is a power series given as 
\begin{equation}
f_0\ =\ 1+\frac3{128}\gamma^2+\frac{15}{16384}\gamma^4+{\cal O}(\gamma^6). 
\end{equation}
The other solution for the exponent $-2$ has a logarithmic part. 
It is given as 
\begin{equation}
g_0\ =\ \gamma^{-2}\left( 1-\frac7{16384}\gamma^4-\frac{21}{1048576}\gamma^6 \right)-\frac1{32}f_0\log\gamma+{\cal O}(\gamma^6). 
\end{equation}
Then, $\pi_0$ is given as 
\begin{eqnarray}
\pi_0 
&=& A_0\gamma+B_0\left( \gamma^2+\frac1{128}\gamma^4+\frac3{16384}\gamma^6 \right)
+C_0\left[ -1+\frac1{32}\gamma^2-\frac1{16384}\gamma^4-\frac3{1048576}\gamma^6 \right. \nonumber \\ [2mm] 
& & \left. -\frac1{32}\left( \gamma^2+\frac1{128}\gamma^4+\frac3{16384}\gamma^6 \right)\log\gamma \right]+{\cal O}(\gamma^8). 
\end{eqnarray}

Since $\pi_0$ is related to the Wilson loop as $\pi_0=-4tW$, this should vanish at $\gamma=0$. 
Therefore, we set $C_0=0$, which eliminates the logarithmic part. 
Then, we obtain 
\begin{eqnarray}
\pi_0 
&=& A_0\gamma+B_0\gamma^2+\frac {B_0}{128}\gamma^4+{\cal O}(\gamma^6), \\ [2mm] 
\pi_1 
&=& 16B_0\gamma+A_0\gamma^2+\frac{3B_0}4\gamma^3+{\cal O}(\gamma^5), \\ [2mm] 
\pi_2 
&=& 4A_0\gamma+28B_0\gamma^2+A_0\gamma^3+\frac{9B_0}{32}\gamma^4+{\cal O}(\gamma^6). 
\end{eqnarray}

From the definition of $\pi_n$, we should have 
\begin{equation}
\pi_0\ \sim\ -4t, \hspace{1cm} \pi_1\ \sim\ -8t, \hspace{1cm} \pi_2\ \sim\ -16t, 
   \label{weak WL-ABJM}
\end{equation}
for small $\gamma$. 
This implies $B_0=A_0/8$. 
The remaining constant $A_0$ will be determined shortly. 

\vspace{5mm}

Next, we consider the second equation (\ref{PF4tH-local}). 
We find that 
\begin{equation}
t\ =\ A_0\left( -\frac14\gamma-\frac1{768}\gamma^3-\frac9{327680}\gamma^5 \right)+{\cal O}(\gamma^7)
   \label{tH-weak-soln}
\end{equation}
is a solution. 
To obtain the general solution of (\ref{PF4tH-local}), we have to solve the following homogeneous equation 
\begin{equation}
4\gamma^2(\gamma-4)(\gamma+4)\partial_\gamma^2\tau+16\gamma(\gamma-2)(\gamma+2)\partial_\gamma\tau+8(\gamma^2+2)\tau\ =\ 0. 
\end{equation}
The integral of $\tau$ can contribute to $t$. 
The characteristic exponents of this equation are $\pm\frac12$. 
The local solutions are 
\begin{eqnarray}
\tau_+ 
&=& \gamma^{\frac12}\left( 1+\frac5{128}\gamma^2+\frac{63}{32768}\gamma^4 \right)+{\cal O}(\gamma^{\frac{13}2}), \\ [2mm] 
\tau_- 
&=& \gamma^{-\frac12}\left( 1+\frac3{128}\gamma^2+\frac{35}{32768}\gamma^4 \right)+{\cal O}(\gamma^{\frac{11}2}). 
\end{eqnarray}
These are not holomorphic at $\gamma=0$. 
If they would contribute to $t$, then the required behaviors (\ref{weak WL-ABJM}) could not be satisfied. 
Therefore, (\ref{tH-weak-soln}) is the solution describing the perturbative limit. 
Note that there was also an additive integration constant which was fixed by requiring $t(0)=0$. 

\vspace{5mm}

The constant $A_0$ was not fixed by the above arguments. 
It is determined by the local behavior of the curve (\ref{curve-ABJM}), as for pure Chern-Simons theory. 
We find that the behavior of $y(x)$ near $z=1$ is 
\begin{equation}
y(z+z^{-1})\ \sim\ -1-\frac1{2(z-1)^2}\gamma
\end{equation}
for small $\gamma$. 
This gives $t\sim\gamma/4$, implying $A_0=-1$. 

\vspace{5mm}

In summary, we have obtained 
\begin{eqnarray}
\pi_0 
&=& -\gamma-\frac18\int d\gamma\,f_0(\gamma)
\ =\ -\gamma-\frac18\gamma^2-\frac1{1024}\gamma^4+{\cal O}(\gamma^6), 
   \label{pi_0-soln}\\ [2mm] 
t 
&=& \frac14\gamma+\frac1{768}\gamma^3+\frac9{327680}\gamma^5+{\cal O}(\gamma^7). 
\end{eqnarray}
Indeed, this reproduces the known result \cite{Drukker:2008zx}\cite{Chen:2008bp}\cite{Rey:2008bh} 
\begin{eqnarray}
W 
&=& e^{t/2}\left( 1-\frac5{24}t^2 \right)+{\cal O}(t^3) \nonumber \\
&=& e^{\pi i\lambda}\left( 1+\frac56\pi^2\lambda^2 \right)+{\cal O}(\lambda^3), \hspace{1cm} \lambda\ =\ \frac Nk, 
\end{eqnarray}
including the framing factor. 
A higher-order extension of \cite{Drukker:2008zx}\cite{Chen:2008bp}\cite{Rey:2008bh} was obtained in \cite{Bianchi:2016yzj}. 

\vspace{5mm}

(ii) $\gamma=\infty$

\vspace{5mm}

Next, we analyze the local solution around $\gamma=\infty$ which describes the gravity dual. 
For this purpose, we introduce a new variable $\xi:=\gamma^{-1}$ and rewrite the equations (\ref{PF4WL-local}) and (\ref{PF4tH-local}) in terms of $\xi$. 
Then, the equations to be solved become 
\begin{equation}
\xi^2(1-16\xi^2)\partial_\xi^2\pi+\xi(16\xi^2-3)\partial_\xi\pi+3\pi\ =\ 0 
   \label{PF4WL-inf}
\end{equation}
and 
\begin{eqnarray}
& & 4\xi^2(1-16\xi^2)\partial_\xi^2\tau-\xi(8+64\xi^2)\partial_\xi\tau+8(1+2\xi^2)\tau \nonumber \\ [2mm] 
&=& \xi(3+16\xi^2)\partial_\xi\pi-(7+32\xi^2)\pi. 
   \label{PF4tH-inf}
\end{eqnarray}
where $\tau=\partial_\gamma t$. 

We start with the equation (\ref{PF4WL-inf}). 
The characteristic exponents at $\xi=0$ turn out to be $1$ and $3$. 
The local solution for the exponent $3$ is 
\begin{equation}
f_\infty\ =\ \xi^3+6\xi^5+60\xi^7+{\cal O}(\xi^9). 
\end{equation}
The other solution for the exponent $1$ is 
\begin{equation}
g_\infty\ =\ \xi-28\xi^5-336\xi^7-8f_\infty\log\xi+{\cal O}(\xi^9). 
\end{equation}
Then, $\pi_0$ is given as 
\begin{eqnarray}
\pi_0 
&=& \frac {A_\infty}\xi+B_\infty\left( -\frac12\xi-\frac32\xi^3-10\xi^5 \right)+C_\infty\left[ -\frac{\log\xi}\xi-2\xi+4\xi^3+\frac{128}3\xi^5 \right. \nonumber \\ [2mm] 
& & \left. +8\left( \frac12\xi+\frac32\xi^3+10\xi^5 \right)\log\xi \right]+{\cal O}(\xi^7\log\xi). 
   \label{pi_0-inf}
\end{eqnarray}
Recall that we have determined the local behavior of $\pi_0$ around $\gamma=0$ completely as (\ref{pi_0-soln}). 
Since the connection formula for the Heun equation is known \cite{Bonelli:2022ten}, the local solution (\ref{pi_0-soln}) can be continued analytically to $\gamma=\infty$. 
This may fix the constants $A_\infty,B_\infty,C_\infty$ in (\ref{pi_0-inf}). 
In the following, we instead determine the coefficients again by using the curve (\ref{curve-ABJM}) as a substitute of the connection formula. 

Next, we consider the equation (\ref{PF4tH-inf}). 
The corresponding homogeneous equation for $\tau$ has the characteristic exponents $1$ and $2$. 
The homogeneous solution for the exponent $2$ is 
\begin{equation}
h(\xi)\ =\ \xi^2+10\xi^4+126\xi^6+{\cal O}(\xi^8). 
\end{equation}
The other homogeneous solution for the exponent $1$ is 
\begin{equation}
\tilde h(\xi)\ =\ \xi+6\xi^3+70\xi^5+924\xi^7+{\cal O}(\xi^9). 
\end{equation}
We find that 
\begin{eqnarray}
i(\xi)\ =\ C_\infty\tilde h(\xi)\log\xi+\frac B4\xi^3+\frac{204B_\infty-664C_\infty}{48}\xi^5+{\cal O}(\xi^7)
\end{eqnarray}
is a solution of the inhomogeneous equation (\ref{PF4tH-inf}). 
Therefore, as long as $C_\infty$ is non-vanishing, $\tau$ behaves as 
\begin{equation}
\tau\ \sim\ C_\infty\xi\log\xi\ =\ -C_\infty\frac{\log\gamma}{\gamma} 
\end{equation}
for small $\xi$ or large $\gamma$. 
Note that the homogeneous solutions $h(\xi)$ and $\tilde h(\xi)$ give only sub-leading terms. 
Since $\tau=\partial_\gamma t$, we find 
\begin{equation}
t\ \sim\ -\frac {C_\infty}2(\log\gamma)^2. 
   \label{tH-strong-ABJM}
\end{equation}
Inverting this relation, we obtain 
\begin{equation}
\gamma\ \sim\ \exp\left( \sqrt{-\frac{2t}{C_\infty}} \right). 
\end{equation}
Then, the expression (\ref{pi_0-inf}) for $\pi_0$ implies  
\begin{equation}
\log \pi_0\ \sim\ \sqrt{-\frac{2t}{C_\infty}}. 
\end{equation}

The constant $C_\infty$ is indeed non-vanishing. 
This can be obtained from the curve (\ref{curve-ABJM}) as 
\begin{equation}
C_\infty\ =\ \frac2{\pi i}. 
\end{equation}
See appendix \ref{integral estimate} for details. 
This reproduces the dual gravity calculation \cite{Drukker:2008zx}\cite{Chen:2008bp}\cite{Rey:2008bh} 
\begin{equation}
\log W\ \sim\ \log\pi_0\ \sim\ \pi\sqrt{2\lambda}. 
\end{equation}

\vspace{5mm}

\subsection{Recursion relation}
   \label{recursion relation ABJM}

\vspace{5mm}

We have obtained a system of differential equations for $\pi_n$ with $n\le3$. 
The other $\pi_n$ may also satisfy suitable differential equations. 
It is much easier to determine them by using a recursion relation, as for pure Chern-Simons theory. 

Recall that we have 
\begin{equation}
2\partial_x(x^ny)\ =\ {\cal S}_n+2^n\frac y{x-2}+(-2)^n\frac y{x+2}-\alpha^n\frac y{x-\alpha}-\beta^n\frac y{x-\beta}, 
\end{equation}
where 
\begin{equation}
{\cal S}_n\ :=\ 2nx^{n-1}y+\frac{x^n-2^n}{x-2}+\frac{x^n-(-2)^n}{x+2}-\frac{x^n-\alpha^n}{x-\alpha}-\frac{x^n-\beta^n}{x-\beta}. 
\end{equation}

We arrange these relations as 
\begin{equation}
\left[
\begin{array}{c}
2\partial_x(x^ny)-{\cal S}_n \\ [1mm]
2\partial_x(x^{n+1}y)-{\cal S}_{n+1} \\ [1mm] 
2\partial_x(x^{n+2}y)-{\cal S}_{n+2} \\ [1mm] 
2\partial_x(x^{n+3}y)-{\cal S}_{n+3}
\end{array}
\right] 
\ =\ 
\left[
\begin{array}{cccc}
2^n & (-2)^n & \alpha^n & \beta^n \\ [1mm] 
2^{n+1} & (-2)^{n+1} & \alpha^{n+1} & \beta^{n+1} \\ [1mm] 
2^{n+2} & (-2)^{n+2} & \alpha^{n+2} & \beta^{n+2} \\ [1mm] 
2^{n+3} & (-2)^{n+3} & \alpha^{n+3} & \beta^{n+3}
\end{array}
\right]
\left[
\begin{array}{c}
\frac y{x-2} \\ [1mm] 
\frac y{x+2} \\ [1mm] 
\frac{-y}{x-\alpha} \\ [1mm] 
\frac{-y}{x-\beta}
\end{array}
\right]. 
\end{equation}
We find an identity 
\begin{eqnarray}
& & (4\alpha\beta,\ -4(\alpha+\beta),\ 4-\alpha\beta,\ \alpha+\beta)
\left[
\begin{array}{cccc}
2^n & (-2)^n & \alpha^n & \beta^n \\
2^{n+1} & (-2)^{n+1} & \alpha^{n+1} & \beta^{n+1} \\
2^{n+2} & (-2)^{n+2} & \alpha^{n+2} & \beta^{n+2} \\
2^{n+3} & (-2)^{n+3} & \alpha^{n+3} & \beta^{n+3}
\end{array}
\right] \nonumber \\
&=& \left( 2^{n+4},\ (-2)^{n+4},\ \alpha^{n+4},\ \beta^{n+4} \right). 
\end{eqnarray}
This can be used to obtain 
\begin{equation}
-4\alpha\beta {\cal S}_n+4(\alpha+\beta){\cal S}_{n+1}+(\alpha\beta-4){\cal S}_{n+2}-(\alpha+\beta){\cal S}_{n+3}\ =\ -S_{n+4}+\partial_x(\cdots). 
\end{equation}
The integration gives a rather simple five-term recursion relation 
\begin{eqnarray}
& & (2n+10)\pi_{n+4}-(2n+9)(\alpha+\beta)\pi_{n+3}+\Bigl[ (2n+8)\alpha\beta-4(2n+4) \Bigr]\pi_{n+2} \nonumber \\ [2mm] 
& &+4(2n+3)(\alpha+\beta)\pi_{n+1}-4(2n+2)\pi_n\ =\ 0. 
\end{eqnarray}
This relation determines all $\pi_n$ with $n\ge4$ in terms of $\pi_0,\cdots,\pi_3$. 

We can show that the generating function $g(t)$ of $\pi_n$ defined as in (\ref{generating fn-pureCS}) satisfies a first-order differential equation which is, however, rather complicated. 
For completeness, we show the equation below: 
\begin{equation}
p\partial_tg+qg\ =\ r, 
\end{equation}
where 
\begin{eqnarray}
p 
&=& 2-2(\alpha+\beta)t+(2\alpha\beta-8)t^2+8(\alpha+\beta)t^3-8\alpha\beta t^4, \\ [2mm] 
q 
&=& -(\alpha+\beta)+(2\alpha\beta+8)t-4(\alpha+\beta)t^2, \\ [2mm] 
r 
&=& 2\pi_0+\Bigl[ 4\pi_1-3(\alpha+\beta)\pi_0 \Bigr]t+\Bigl[ 6\pi_2-5(\alpha+\beta)\pi_1+4\alpha\beta\pi_0 \Bigr]t^2. 
\end{eqnarray}

\vspace{1cm}

\section{Discussion}
   \label{Discussion}

\vspace{5mm}

We have derived the Picard-Fuchs equations from the curves (\ref{curve-pureCS}) and (\ref{curve-ABJM}) for pure Chern-Simons theory and ABJM theory, respectively, which are satisfied by the BPS Wilson loops and the 't~Hooft coupling. 
The curves are obtained from the derivative of the resolvents of the theories which have simpler forms compared with the resolvents themselves, and therefore, the analysis becomes simpler. 
We have checked that the Picard-Fuchs equations give the known results correctly. 
We have also obtained recursion relations for the quantities $\pi_n$ from which we can obtain the planar vevs of Wilson loops in an arbitrary representation. 

Since our analysis does not rely on any relation to (topological) string theory, we expect that it can be extended to more general Chern-Simons-matter theories, irrespective of whether there exist gravity duals for them. 
The exploration of various theories may help us to characterize the family of Chern-Simons-matter theories with gravity duals. 

An immediate extension of our analysis will be to consider ABJ theory. 
This requires us to vary two parameters $\alpha$ and $\beta$ independently. 
This results in Picard-Fuchs equations which are a system of partial differential equations. 
In this context, we are interested in the following two issues. 
The one is the conjecture in \cite{Aharony:2008gk} which claims that the theory does not exist as a unitary theory when the 't~Hooft couplings does not satisfy a certain condition. 
It is interesting to see whether and how it appears in the Picard-Fuchs equations. 
The other one is the strong coupling limit. 
In \cite{Marino:2009jd}, it is observed that the strong coupling behavior of the Wilson loops may depend of how we take the strong coupling limit. 
This should be reflected in the local behavior of the Picard-Fuchs equations. 

It is known \cite{Giveon:2008zn} that a similar phenomenon occurs in Chern-Simons theory with flavors in which the supersymmetry is broken in some parameter region. 
This case is much simpler than ABJ theory, and the analysis of this theory will be helpful to understand ABJ theory. 
It will be also interesting to reconsider ABJM theory with flavors, which is discussed in \cite{Hikida:2009tp}\cite{Santamaria:2010dm}, in light of our analysis. 

Another interesting extension will be to consider GT theory. 
In this theory, it is known \cite{Suyama:2011yz} that the strong coupling behavior of the Wilson loop is qualitatively different from the one in ABJM theory. 
It is interesting to clarify how the difference arises. 
One more direction of the extension is to change the number of bi-fundamental matters. 
Such theories are discussed in \cite{Suyama:2016nap}. 
Comparing these extensions, we hope to understand how information on various physically interesting limits is
encoded in the analytic properties of the Picard-Fuchs equations. 

\appendix

\vspace{2cm}

\section{Derivation of the curves}
   \label{curve-derivation}

\vspace{5mm}

In this appendix, we summarize how to derive the curves (\ref{curve-pureCS}) and (\ref{curve-ABJM}) defined by the derivative $zv'(z)$ of the resolvent $v(z)$, without determining $v(z)$ itself. 
In fact, $zv'(z)$ has a simpler structure than $v(z)$.  

First, we consider the matrix model for pure Chern-Simons theory to illustrate the derivation. 
The saddle point equations (\ref{saddle-pureCS}) can be rewritten in terms of the resolvent (\ref{resolvent-pureCS}) as 
\begin{equation}
2\log x\ =\ v(x_+)+v(x_-), \hspace{1cm} x\in I, 
\end{equation}
where $I$ is the branch cut of $v(z)$. 
This equation can be simplified by differentiating both sides by $x$. 
As a result, we obtain 
\begin{equation}
\omega(x_+)+\omega(x_-)\ =\ 0, \hspace{1cm} \omega(z)\ :=\ zv'(z)-1. 
\end{equation}
This implies that $\omega(z)$ changes its sign when it passes the branch cut $I$. 
Due to the Wigner law for $v(z)$, $\omega(z)$ behaves as 
\begin{equation}
\omega(z)\ \sim\ \frac{c}{\sqrt{z-a}}
\end{equation}
around $z=a$, and similarly for $z=a^{-1}$. 
The definition (\ref{resolvent-pureCS}) of $v(z)$ implies $\omega(\infty)=-1$. 

From these analytic properties, we conclude that $\omega(z)^2$ is a rational function on $\mathbb{C}$ with simple poles at $z=a,a^{-1}$. 
The asymptotic value of $\omega(z)$ implies 
\begin{equation}
\omega(z)^2\ =\ \frac{z^2+Az+B}{(z-a)(z-a^{-1})}. 
\end{equation}
We have the inversion symmetry $\omega(z^{-1})=\omega(z)$ followed from the symmetry of the saddle point equations (\ref{saddle-pureCS}). 
This implies $B=1$. 
If the numerator of the right-hand side is not a complete square, then $\omega(z)$ would have extra branch points
other than $z=a,a^{-1}$. 
We are interested in the single-cut solution, so we should choose $A=\pm2$. 

To fix the sign of $A$, recall that the non-analytic part of $v(z)$ should be proportional to $\sqrt{(z-a)(z-a^{-1})}$. 
In the limit $a\to1$, this becomes $z-1$, and therefore $\omega(z)$ should not have any pole at $z=1$ in the limit. 
This can be achieved only when we choose $A=-2$. 
Now we conclude that $\omega(z)$ satisfies 
\begin{equation}
\omega(z)^2\ =\ \frac{(z-1)^2}{(z-a)(z-a^{-1})}, 
\end{equation}
which reproduces the curve (\ref{curve-pureCS}). 

The curve (\ref{curve-ABJM}) for ABJM theory can be derived similarly. 
The resolvents $v_1(z)$ and $v_2(z)$ defined in (\ref{resolvent1-ABJM})(\ref{resolvent2-ABJM}) satisfy 
\begin{eqnarray}
2\log x\ =\ v_1(x_+)+v_2(x_-)-2v_2(x), \hspace{1cm} x\in I_1 \\ [2mm] 
-2\log(-\tilde x)\ =\ v_2(\tilde x_+)+v_2(\tilde x_-)-2v_1(\tilde x). \hspace{1cm} \tilde x\in I_2
\end{eqnarray}
By the differentiation, we obtain 
\begin{equation}
\omega(x_+)+\omega(x_-)\ =\ 0, \hspace{1cm} \omega(\tilde x_+)+\omega(\tilde x_-)\ =\ 0, 
\end{equation}
where 
\begin{equation}
\omega(z)\ :=\ zv_1'(z)-zv_2'(z)-1. 
\end{equation}
These equations, the Wigner law, the asymptotic value, and the absence of extra branch points imply 
\begin{equation}
\omega(z)^2\ =\ \frac{(z^2+Az+B)^2}{(z-a)(z-a^{-1})(z-b)(z-b^{-1})}. 
\end{equation}
The inversion symmetry of $\omega(z)$ requires 
\begin{equation}
B\ =\ \pm1, \hspace{1cm} A\ =\ AB. 
\end{equation}
By the same argument as for pure Chern-Simons theory given above, $\omega(z)^2$ must be holomorphic in the limit $a\to1$ and $b\to-1$ simultaneously. 
This can be realized only when we choose $A=0$ and $B=-1$. 
We conclude that $\omega(z)$ satisfies 
\begin{equation}
\omega(z)^2\ =\ \frac{(z^2-1)^2}{(z-a)(z-a^{-1})(z-b)(z-b^{-1})}. 
\end{equation}
This is the curve (\ref{curve-ABJM}) for ABJM theory. 

\vspace{1cm}

\section{Strong coupling behavior}
   \label{integral estimate}

\vspace{5mm}

In this appendix, we calculate the leading term of the 't~Hooft coupling of ABJM theory in the strong coupling limit. 
This corresponds to extracting a divergent contribution in the limit from the integral expression 
\begin{equation}
t\ =\ -\frac1{4\pi i}\int_C\frac{dz}z\log z\,y(z+z^{-1}), 
\end{equation}
where $y(x)$ defines the curve (\ref{curve-ABJM}). 

\vspace{5mm}

It is instructive to perform a similar calculation for pure Chern-Simons theory. 
The 't~Hooft coupling is given as 
\begin{equation}
t\ =\ \frac1{4\pi i}\int_C\frac{dz}z\log z\,\frac{z-1}{\sqrt{(z-a)(z-a^{-1})}}. 
\end{equation}
The integration contour $C$ is depicted in Figure \ref{fig-pureCS}. 
We are interested in a limit $\alpha\to\infty$ which was found as one of the singularities of the Picard-Fuchs equation (\ref{PF-pi_0}). 
This limit corresponds to $a\to0$. 

\begin{figure}
\begin{center}
\includegraphics{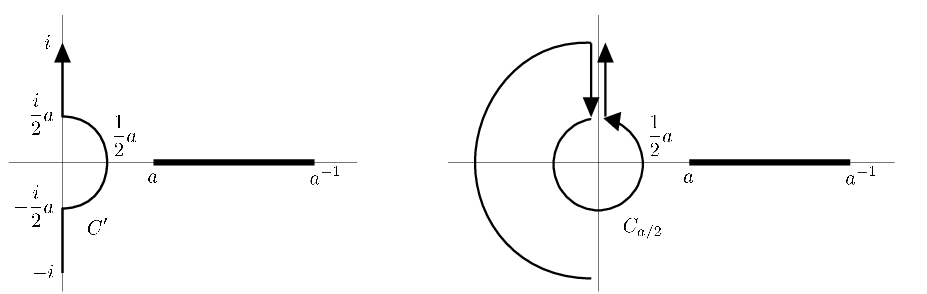}
\end{center}
\caption{
Deformations of the integration contour. 
The integration along the contour $C'$ is $-\frac12$ times the original integral. 
The contour in the right figure is a deformation of $C'$. 
The leading contribution comes from the circular part $C_{a/2}$. 
}
   \label{fig3-pureCS}
\end{figure}

In the limit, the contour is pinched by the branch point at $z=a$ and the singularity at $z=0$, resulting in a divergent behavior. 
Since the integral expression is invariant under $z\to z^{-1}$, we see that another pinching occurs at $z=\infty$. 
To simplify the calculation, we use this symmetry to rewrite the above integral as 
\begin{equation}
t\ =\ -\frac1{2\pi i}\int_{C'}\frac{dz}z\log z\,\frac{z-1}{\sqrt{(z-a)(z-a^{-1})}}, 
\end{equation}
where the contour $C'$ is depicted in the left of Figure \ref{fig3-pureCS}. 

We are interested in the leading term for small $a$. 
In order to calculate this term, we deform the contour $C'$ as in the right of Figure \ref{fig3-pureCS}. 
Then we find that only the part $C_{a/2}$ gives a divergent behavior for $t$. 
By rescaling the variable $z$, the corresponding integral can be estimated as 
\begin{eqnarray}
t 
&\sim& -\frac1{2\pi i}\int_{C_1}\frac{dz}z\log a\frac{az-1}{\sqrt{(az-a)(az-a^{-1})}}
\ =\ -\log a\ \sim\ \log\alpha, 
\end{eqnarray}
where $C_1$ is the unit circle with the counter-clockwise direction. 
This indeed reproduces the large $\alpha$ behavior of $t$ derived from the exact result (\ref{solution-pureCS}). 
The point of this calculation is that the pinching of the integration contour gives a divergent factor $\log\alpha$. 

\vspace{5mm}

Now, we consider the 't~Hooft coupling 
\begin{equation}
t\ =\ \frac1{4\pi i}\int_C\frac{dz}z\log z\frac{z^2-1}{\sqrt{(z-a)(z-a^{-1})(z-b)(z-b^{-1})}}
\end{equation}
of ABJM theory in the limit $\gamma\to\infty$. 
We choose $C_1$ in Figure \ref{fig-ABJM} for the contour $C$. 
Recall that $\gamma$ determines $a$ and $b$ by 
\begin{equation}
a+a^{-1}\ =\ \alpha\ =\ 2+\gamma, \hspace{1cm} b+b^{-1}\ =\ \beta\ =\ -2+\gamma 
\end{equation}
on the ABJM slice. 
For large $\gamma$, this implies 
\begin{equation}
a\ \sim\ \frac1\gamma z_-, \hspace{1cm} b\ \sim\ \frac1\gamma z_+, \hspace{1cm} z_\pm\ :=\ 1\pm\frac2\gamma. 
\end{equation}
Note that both $a$ and $b$ approach the origin where the integrand is singular. 

\begin{figure}
\begin{center}
\includegraphics{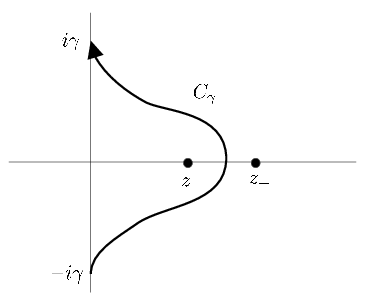}
\end{center}
\caption{
The contour $C_\gamma$ is schematically depicted. 
The leading contribution comes from the part of $C_\gamma$ near $z=z_\pm$, so the details of the contour and the branch cuts are not relevant. 
}
   \label{fig2-ABJM}
\end{figure}

As for pure Chern-Simons theory, $t$ can be rewritten as 
\begin{equation}
t\ =\ -\frac1{2\pi i}\int_{C'}\frac{dz}z\log z\frac{z^2-1}{\sqrt{(z-a)(z-a^{-1})(z-b)(z-b^{-1})}}
\end{equation}
for the contour $C'$ depicted in Figure \ref{fig3-pureCS}. 
The divergent behavior then appears from the pinching at $z=0$. 
By rescaling the variable $z$, we find 
\begin{equation}
t\ \sim\ -\frac1{2\pi i}\int_{C_\gamma}\frac{dz}z\log\frac1\gamma\frac{z^2/\gamma^2-1}{\sqrt{(z-z_-)(z/\gamma^2-z_-^{-1})(z-z_+)(z/\gamma^2-z_+^{-1})}} 
\end{equation}
for large $\gamma$. 
The contour $C_\gamma$ is depicted in Figure \ref{fig2-ABJM}. 
Note that the pinching at $z=0$ in the original expression is replaced with a divergent factor $\log\gamma$. 

Interestingly, there is another source of divergence. 
The contour $C_\gamma$ is pinched by the branch points at $z=z_\pm$ both of which approach $z=1$. 
The divergent part can be estimated as 
\begin{equation}
t\ \sim\ -\frac1{2\pi i}\log\gamma\int_{1-i}^{1+i}\frac{dz}{\sqrt{(z-z_-)(z-z_+)}}\ \sim\ -\frac1{\pi i}(\log\gamma^2). 
\end{equation}
This implies that one of the integration constants appearing in (\ref{tH-strong-ABJM}) turns out to be 
\begin{equation}
C_\infty\ =\ \frac2{\pi i}. 
\end{equation}


\vspace{1cm}

\begin{thebibliography}{99}

%\cite{Aharony:2008ug}
\bibitem{Aharony:2008ug}
O.~Aharony, O.~Bergman, D.~L.~Jafferis and J.~Maldacena,
``N=6 superconformal Chern-Simons-matter theories, M2-branes and their gravity duals,''
JHEP \textbf{10}, 091 (2008)
doi:10.1088/1126-6708/2008/10/091
[arXiv:0806.1218 [hep-th]].
%2553 citations counted in INSPIRE as of 01 Sep 2025

%\cite{Maldacena:1997re}
\bibitem{Maldacena:1997re}
J.~M.~Maldacena,
``The Large $N$ limit of superconformal field theories and supergravity,''
Adv. Theor. Math. Phys. \textbf{2}, 231-252 (1998)
doi:10.4310/ATMP.1998.v2.n2.a1
[arXiv:hep-th/9711200 [hep-th]].
%21146 citations counted in INSPIRE as of 01 Sep 2025

%\cite{Aharony:2008gk}
\bibitem{Aharony:2008gk}
O.~Aharony, O.~Bergman and D.~L.~Jafferis,
``Fractional M2-branes,''
JHEP \textbf{11}, 043 (2008)
doi:10.1088/1126-6708/2008/11/043
[arXiv:0807.4924 [hep-th]].
%682 citations counted in INSPIRE as of 01 Sep 2025

%\cite{Hosomichi:2008jb}
\bibitem{Hosomichi:2008jb}
K.~Hosomichi, K.~M.~Lee, S.~Lee, S.~Lee and J.~Park,
``N=5,6 Superconformal Chern-Simons Theories and M2-branes on Orbifolds,''
JHEP \textbf{09}, 002 (2008)
doi:10.1088/1126-6708/2008/09/002
[arXiv:0806.4977 [hep-th]].
%376 citations counted in INSPIRE as of 01 Sep 2025

%\cite{Gaiotto:2008sd}
\bibitem{Gaiotto:2008sd}
D.~Gaiotto and E.~Witten,
``Janus Configurations, Chern-Simons Couplings, And The theta-Angle in N=4 Super Yang-Mills Theory,''
JHEP \textbf{06}, 097 (2010)
doi:10.1007/JHEP06(2010)097
[arXiv:0804.2907 [hep-th]].
%456 citations counted in INSPIRE as of 01 Sep 2025

%\cite{Drukker:2008zx}
\bibitem{Drukker:2008zx}
N.~Drukker, J.~Plefka and D.~Young,
``Wilson loops in 3-dimensional N=6 supersymmetric Chern-Simons Theory and their string theory duals,''
JHEP \textbf{11}, 019 (2008)
doi:10.1088/1126-6708/2008/11/019
[arXiv:0809.2787 [hep-th]].
%189 citations counted in INSPIRE as of 01 Sep 2025

%\cite{Chen:2008bp}
\bibitem{Chen:2008bp}
B.~Chen and J.~B.~Wu,
``Supersymmetric Wilson Loops in N=6 Super Chern-Simons-matter theory,''
Nucl. Phys. B \textbf{825}, 38-51 (2010)
doi:10.1016/j.nuclphysb.2009.09.015
[arXiv:0809.2863 [hep-th]].
%158 citations counted in INSPIRE as of 01 Sep 2025

%\cite{Rey:2008bh}
\bibitem{Rey:2008bh}
S.~J.~Rey, T.~Suyama and S.~Yamaguchi,
``Wilson Loops in Superconformal Chern-Simons Theory and Fundamental Strings in Anti-de Sitter Supergravity Dual,''
JHEP \textbf{03}, 127 (2009)
doi:10.1088/1126-6708/2009/03/127
[arXiv:0809.3786 [hep-th]].
%152 citations counted in INSPIRE as of 01 Sep 2025

%\cite{Rey:1998ik}
\bibitem{Rey:1998ik}
S.~J.~Rey and J.~T.~Yee,
``Macroscopic strings as heavy quarks in large N gauge theory and anti-de Sitter supergravity,''
Eur. Phys. J. C \textbf{22}, 379-394 (2001)
doi:10.1007/s100520100799
[arXiv:hep-th/9803001 [hep-th]].
%1510 citations counted in INSPIRE as of 01 Sep 2025

%\cite{Maldacena:1998im}
\bibitem{Maldacena:1998im}
J.~M.~Maldacena,
``Wilson loops in large N field theories,''
Phys. Rev. Lett. \textbf{80}, 4859-4862 (1998)
doi:10.1103/PhysRevLett.80.4859
[arXiv:hep-th/9803002 [hep-th]].
%2092 citations counted in INSPIRE as of 01 Sep 2025

%\cite{Bianchi:2016yzj}
\bibitem{Bianchi:2016yzj}
M.~S.~Bianchi, L.~Griguolo, M.~Leoni, A.~Mauri, S.~Penati and D.~Seminara,
``Framing and localization in Chern-Simons theories with matter,''
JHEP \textbf{06}, 133 (2016)
doi:10.1007/JHEP06(2016)133
[arXiv:1604.00383 [hep-th]].
%30 citations counted in INSPIRE as of 02 Sep 2025

%\cite{Drukker:2009hy}
\bibitem{Drukker:2009hy}
N.~Drukker and D.~Trancanelli,
``A Supermatrix model for N=6 super Chern-Simons-matter theory,''
JHEP \textbf{02}, 058 (2010)
doi:10.1007/JHEP02(2010)058
[arXiv:0912.3006 [hep-th]].
%206 citations counted in INSPIRE as of 02 Sep 2025

%\cite{Lee:2010hk}
\bibitem{Lee:2010hk}
K.~M.~Lee and S.~Lee,
``1/2-BPS Wilson Loops and Vortices in ABJM Model,''
JHEP \textbf{09}, 004 (2010)
doi:10.1007/JHEP09(2010)004
[arXiv:1006.5589 [hep-th]].
%61 citations counted in INSPIRE as of 02 Sep 2025

%\cite{Drukker:2019bev}
\bibitem{Drukker:2019bev}
N.~Drukker, D.~Trancanelli, L.~Bianchi, M.~S.~Bianchi, D.~H.~Correa, V.~Forini, L.~Griguolo, M.~Leoni, F.~Levkovich-Maslyuk and G.~Nagaoka, \textit{et al.}
``Roadmap on Wilson loops in 3d Chern{\textendash}Simons-matter theories,''
J. Phys. A \textbf{53}, no.17, 173001 (2020)
doi:10.1088/1751-8121/ab5d50
[arXiv:1910.00588 [hep-th]].
%78 citations counted in INSPIRE as of 02 Sep 2025

%\cite{Kapustin:2009kz}
\bibitem{Kapustin:2009kz}
A.~Kapustin, B.~Willett and I.~Yaakov,
``Exact Results for Wilson Loops in Superconformal Chern-Simons Theories with Matter,''
JHEP \textbf{03}, 089 (2010)
doi:10.1007/JHEP03(2010)089
[arXiv:0909.4559 [hep-th]].
%919 citations counted in INSPIRE as of 02 Sep 2025

%\cite{Suyama:2009pd}
\bibitem{Suyama:2009pd}
T.~Suyama,
``On Large N Solution of ABJM Theory,''
Nucl. Phys. B \textbf{834}, 50-76 (2010)
doi:10.1016/j.nuclphysb.2010.03.011
[arXiv:0912.1084 [hep-th]].
%36 citations counted in INSPIRE as of 02 Sep 2025

%\cite{Marino:2009jd}
\bibitem{Marino:2009jd}
M.~Marino and P.~Putrov,
``Exact Results in ABJM Theory from Topological Strings,''
JHEP \textbf{06}, 011 (2010)
doi:10.1007/JHEP06(2010)011
[arXiv:0912.3074 [hep-th]].
%244 citations counted in INSPIRE as of 02 Sep 2025

%\cite{Drukker:2010nc}
\bibitem{Drukker:2010nc}
N.~Drukker, M.~Marino and P.~Putrov,
``From weak to strong coupling in ABJM theory,''
Commun. Math. Phys. \textbf{306}, 511-563 (2011)
doi:10.1007/s00220-011-1253-6
[arXiv:1007.3837 [hep-th]].
%500 citations counted in INSPIRE as of 02 Sep 2025

%\cite{Halmagyi:2003ze}
\bibitem{Halmagyi:2003ze}
N.~Halmagyi and V.~Yasnov,
``The Spectral curve of the lens space matrix model,''
JHEP \textbf{11}, 104 (2009)
doi:10.1088/1126-6708/2009/11/104
[arXiv:hep-th/0311117 [hep-th]].
%82 citations counted in INSPIRE as of 02 Sep 2025

%\cite{Aguilera-Damia:2014qgy}
\bibitem{Aguilera-Damia:2014qgy}
J.~Aguilera-Damia, D.~H.~Correa and G.~A.~Silva,
``Strings in $AdS_4 \times \mathbb{CP}^{3}$ Wilson loops in $\mathcal N=$6 super Chern-Simons-matter and bremsstrahlung functions,''
JHEP \textbf{06}, 139 (2014)
doi:10.1007/JHEP06(2014)139
[arXiv:1405.1396 [hep-th]].
%51 citations counted in INSPIRE as of 02 Sep 2025

%\cite{Marino:2011eh}
\bibitem{Marino:2011eh}
M.~Marino and P.~Putrov,
``ABJM theory as a Fermi gas,''
J. Stat. Mech. \textbf{1203}, P03001 (2012)
doi:10.1088/1742-5468/2012/03/P03001
[arXiv:1110.4066 [hep-th]].
%326 citations counted in INSPIRE as of 02 Sep 2025

%\cite{Hatsuda:2015gca}
\bibitem{Hatsuda:2015gca}
Y.~Hatsuda, S.~Moriyama and K.~Okuyama,
``Exact instanton expansion of the ABJM partition function,''
PTEP \textbf{2015}, no.11, 11B104 (2015)
doi:10.1093/ptep/ptv145
[arXiv:1507.01678 [hep-th]].
%34 citations counted in INSPIRE as of 02 Sep 2025

%\cite{Marino:2016new}
\bibitem{Marino:2016new}
M.~Marino,
``Localization at large N in Chern{\textendash}Simons-matter theories,''
J. Phys. A \textbf{50}, no.44, 443007 (2017)
doi:10.1088/1751-8121/aa5f69
[arXiv:1608.02959 [hep-th]].
%37 citations counted in INSPIRE as of 02 Sep 2025

%\cite{Klemm:2012ii}
\bibitem{Klemm:2012ii}
A.~Klemm, M.~Marino, M.~Schiereck and M.~Soroush,
``Aharony{\textendash}Bergman{\textendash}Jafferis{\textendash}Maldacena Wilson Loops in the Fermi Gas Approach,''
Z. Naturforsch. A \textbf{68}, no.1-2, 178-209 (2013)
doi:10.5560/zna.2012-0118
[arXiv:1207.0611 [hep-th]].
%101 citations counted in INSPIRE as of 02 Sep 2025

%\cite{Hatsuda:2013yua}
\bibitem{Hatsuda:2013yua}
Y.~Hatsuda, M.~Honda, S.~Moriyama and K.~Okuyama,
``ABJM Wilson Loops in Arbitrary Representations,''
JHEP \textbf{10}, 168 (2013)
doi:10.1007/JHEP10(2013)168
[arXiv:1306.4297 [hep-th]].
%60 citations counted in INSPIRE as of 02 Sep 2025

%\cite{Grassi:2014zfa}
\bibitem{Grassi:2014zfa}
A.~Grassi, Y.~Hatsuda and M.~Marino,
``Topological Strings from Quantum Mechanics,''
Annales Henri Poincare \textbf{17}, no.11, 3177-3235 (2016)
doi:10.1007/s00023-016-0479-4
[arXiv:1410.3382 [hep-th]].
%198 citations counted in INSPIRE as of 03 Sep 2025

%\cite{Marino:2015nla}
\bibitem{Marino:2015nla}
M.~Marino,
``Spectral theory and mirror symmetry.,''
Proc. Symp. Pure Math. \textbf{98}, 259 (2018)
doi:10.1090/pspum/098/01722
[arXiv:1506.07757 [math-ph]].
%68 citations counted in INSPIRE as of 03 Sep 2025

%\cite{Suyama:2016nap}
\bibitem{Suyama:2016nap}
T.~Suyama,
``Notes on Planar Resolvents of Chern-Simons-matter Matrix Models,''
JHEP \textbf{11}, 049 (2016)
doi:10.1007/JHEP11(2016)049
[arXiv:1605.09110 [hep-th]].
%6 citations counted in INSPIRE as of 08 Sep 2025

%\cite{Isidro:2000zw}
\bibitem{Isidro:2000zw}
J.~M.~Isidro,
``Integrability, Seiberg-Witten models and Picard-Fuchs equations,''
JHEP \textbf{01}, 043 (2001)
doi:10.1088/1126-6708/2001/01/043
[arXiv:hep-th/0011253 [hep-th]].
%11 citations counted in INSPIRE as of 15 Sep 2025

%\cite{Witten:1988hf}
\bibitem{Witten:1988hf}
E.~Witten,
``Quantum Field Theory and the Jones Polynomial,''
Commun. Math. Phys. \textbf{121}, 351-399 (1989)
doi:10.1007/BF01217730
%3927 citations counted in INSPIRE as of 28 Aug 2025

%\cite{Marino:2002fk}
\bibitem{Marino:2002fk}
M.~Marino,
``Chern-Simons theory, matrix integrals, and perturbative three manifold invariants,''
Commun. Math. Phys. \textbf{253}, 25-49 (2004)
doi:10.1007/s00220-004-1194-4
[arXiv:hep-th/0207096 [hep-th]].
%252 citations counted in INSPIRE as of 28 Aug 2025

%\cite{Aganagic:2002wv}
\bibitem{Aganagic:2002wv}
M.~Aganagic, A.~Klemm, M.~Marino and C.~Vafa,
``Matrix model as a mirror of Chern-Simons theory,''
JHEP \textbf{02}, 010 (2004)
doi:10.1088/1126-6708/2004/02/010
[arXiv:hep-th/0211098 [hep-th]].
%285 citations counted in INSPIRE as of 28 Aug 2025

%\cite{Witten:1992fb}
\bibitem{Witten:1992fb}
E.~Witten,
``Chern-Simons gauge theory as a string theory,''
Prog. Math. \textbf{133}, 637-678 (1995)
[arXiv:hep-th/9207094 [hep-th]].
%638 citations counted in INSPIRE as of 29 Aug 2025

%\cite{Dijkgraaf:2002fc}
\bibitem{Dijkgraaf:2002fc}
R.~Dijkgraaf and C.~Vafa,
``Matrix models, topological strings, and supersymmetric gauge theories,''
Nucl. Phys. B \textbf{644}, 3-20 (2002)
doi:10.1016/S0550-3213(02)00766-6
[arXiv:hep-th/0206255 [hep-th]].
%657 citations counted in INSPIRE as of 29 Aug 2025

%\cite{Gopakumar:1998ki}
\bibitem{Gopakumar:1998ki}
R.~Gopakumar and C.~Vafa,
``On the gauge theory / geometry correspondence,''
Adv. Theor. Math. Phys. \textbf{3}, 1415-1443 (1999)
doi:10.4310/ATMP.1999.v3.n5.a5
[arXiv:hep-th/9811131 [hep-th]].
%730 citations counted in INSPIRE as of 29 Aug 2025

%%\cite{Kapustin:2009kz}
%\bibitem{Kapustin:2009kz}
%A.~Kapustin, B.~Willett and I.~Yaakov,
%``Exact Results for Wilson Loops in Superconformal Chern-Simons Theories with Matter,''
%JHEP \textbf{03}, 089 (2010)
%doi:10.1007/JHEP03(2010)089
%[arXiv:0909.4559 [hep-th]].
%%917 citations counted in INSPIRE as of 29 Aug 2025

%\cite{Kao:1995gf}
\bibitem{Kao:1995gf}
H.~C.~Kao, K.~M.~Lee and T.~Lee,
``The Chern-Simons coefficient in supersymmetric Yang-Mills Chern-Simons theories,''
Phys. Lett. B \textbf{373}, 94-99 (1996)
doi:10.1016/0370-2693(96)00119-0
[arXiv:hep-th/9506170 [hep-th]].
%106 citations counted in INSPIRE as of 28 Aug 2025

%\cite{Morita:2017oev}
\bibitem{Morita:2017oev}
T.~Morita and K.~Sugiyama,
``Multi-cut solutions in Chern{\textendash}Simons matrix models,''
Nucl. Phys. B \textbf{929}, 1-20 (2018)
doi:10.1016/j.nuclphysb.2018.01.028
[arXiv:1704.08675 [hep-th]].
%8 citations counted in INSPIRE as of 08 Sep 2025

%\cite{Morita:2018oel}
\bibitem{Morita:2018oel}
T.~Morita and K.~Sugiyama,
``Toward the construction of the general multi-cut solutions in Chern-Simons Matrix Models,''
JHEP \textbf{08}, 168 (2018)
doi:10.1007/JHEP08(2018)168
[arXiv:1805.00831 [hep-th]].
%2 citations counted in INSPIRE as of 08 Sep 2025

%\cite{Gross:1998gk}
\bibitem{Gross:1998gk}
D.~J.~Gross and H.~Ooguri,
``Aspects of large N gauge theory dynamics as seen by string theory,''
Phys. Rev. D \textbf{58}, 106002 (1998)
doi:10.1103/PhysRevD.58.106002
[arXiv:hep-th/9805129 [hep-th]].
%419 citations counted in INSPIRE as of 08 Sep 2025

%\cite{Gaiotto:2009mv}
\bibitem{Gaiotto:2009mv}
D.~Gaiotto and A.~Tomasiello,
``The gauge dual of Romans mass,''
JHEP \textbf{01}, 015 (2010)
doi:10.1007/JHEP01(2010)015
[arXiv:0901.0969 [hep-th]].
%152 citations counted in INSPIRE as of 09 Sep 2025

\bibitem{Heun}
K.~Takemura. ``On the Heun Equation''. 
In: Philosophical Transactions: Mathematical, Physical and Engineering Sciences 366.1867 (2008), pp. 1179–1201. 
issn: 1364503X. url: http://www.jstor.org/stable/25190740.

%\cite{Bonelli:2022ten}
\bibitem{Bonelli:2022ten}
G.~Bonelli, C.~Iossa, D.~Panea Lichtig and A.~Tanzini,
``Irregular Liouville Correlators and Connection Formulae for Heun Functions,''
Commun. Math. Phys. \textbf{397}, no.2, 635-727 (2023)
doi:10.1007/s00220-022-04497-5
[arXiv:2201.04491 [hep-th]].
%105 citations counted in INSPIRE as of 15 Sep 2025

%\cite{Giveon:2008zn}
\bibitem{Giveon:2008zn}
A.~Giveon and D.~Kutasov,
``Seiberg Duality in Chern-Simons Theory,''
Nucl. Phys. B \textbf{812}, 1-11 (2009)
doi:10.1016/j.nuclphysb.2008.09.045
[arXiv:0808.0360 [hep-th]].
%262 citations counted in INSPIRE as of 12 Sep 2025

%\cite{Hikida:2009tp}
\bibitem{Hikida:2009tp}
Y.~Hikida, W.~Li and T.~Takayanagi,
``ABJM with Flavors and FQHE,''
JHEP \textbf{07}, 065 (2009)
doi:10.1088/1126-6708/2009/07/065
[arXiv:0903.2194 [hep-th]].
%119 citations counted in INSPIRE as of 12 Sep 2025

%\cite{Santamaria:2010dm}
\bibitem{Santamaria:2010dm}
R.~C.~Santamaria, M.~Marino and P.~Putrov,
``Unquenched flavor and tropical geometry in strongly coupled Chern-Simons-matter theories,''
JHEP \textbf{10}, 139 (2011)
doi:10.1007/JHEP10(2011)139
[arXiv:1011.6281 [hep-th]].
%54 citations counted in INSPIRE as of 11 Sep 2025

%\cite{Suyama:2011yz}
\bibitem{Suyama:2011yz}
T.~Suyama,
``Eigenvalue Distributions in Matrix Models for Chern-Simons-matter Theories,''
Nucl. Phys. B \textbf{856}, 497-527 (2012)
doi:10.1016/j.nuclphysb.2011.11.013
[arXiv:1106.3147 [hep-th]].
%23 citations counted in INSPIRE as of 12 Sep 2025

\end{thebibliography}
\end{document}